\documentclass[journal,11pt,draftclsnofoot,onecolumn]{IEEEtran}

\usepackage{graphicx}
\usepackage[square,comma,numbers,sort&compress]{natbib}
\usepackage{amssymb,amscd,amssymb,amsmath,mathrsfs}
\usepackage{epstopdf}
\usepackage{balance}
\usepackage[noend]{algorithm} 
\usepackage{algorithmicx} 
\usepackage[noend]{algpseudocode}
\usepackage{multirow} 
\usepackage{amsmath}
\usepackage{xcolor}
\usepackage{subfigure}
\usepackage{setspace}
\usepackage{enumerate}
\usepackage{lscape}
\usepackage{framed}
\usepackage[noend]{algpseudocode}
\usepackage{algorithmicx,algorithm}
\floatname{algorithm}{Algorithm}
\usepackage{etoolbox}

\makeatletter

\newcommand{\Rmnum}[1]{\expandafter\@slowromancap\romannumeral #1@}
\makeatother

\newtheorem{thm}{Theorem}
\newtheorem{pro}{Proposition}

\newtheorem{defn}{Definition}
\newtheorem{example}{Example}

\newtheorem{remark}{Remark}

\begin{document}
%
\title{\LARGE \bf
Bounded-Time Nonblocking Supervisory Control of Timed Discrete-Event Systems}

\author{Renyuan Zhang$^{1}$, Jiale Wu$^{1}$, Junhua Gou$^{1}$, Yabo Zhu$^{1}$ and Kai Cai$^{2}$
\thanks{*This work was supported in part by the Natural Science Foundation of Shaanxi Province, China, Grant no. 2021JM-072.}
\thanks{$^{1}$R. Zhang, J. Wu, J. Gou and Y. Zhu are with the School of Automation, Northwestern Polytechnical University, Xi'an, China
        ({\tt\small ryzhang@nwpu.edu.cn}).}%
\thanks{$^{2}$K. Cai is with the Control and Artificial Intelligence Group, Department of Core Informatics, Osaka Metropolitan University, Osaka, Japan
        ({\tt\small kai.cai@eng.osaka-cu.ac.jp}).}%
}

\maketitle

\thispagestyle{empty} \pagestyle{plain}

\begin{abstract}
Recently an automaton property of quantitative nonblockingness was proposed in supervisory control of untimed discrete-event systems (DES), which {\em quantifies} the standard nonblocking property by capturing the practical requirement that all tasks be completed within a bounded number of steps. However, in practice tasks may be further required to be completed in specific time. To meet this new requirement, in this paper we introduce the concept of {\em bounded-time nonblockingness}, which extends the concept of
quantitative nonblockingness from untimed DES to timed DES. This property requires that each task must be completed within a bounded time counted by the number of $ticks$, rather than
bounded number of transition steps in quantitative nonblockingness. Accordingly, we formulate a new bounded-time nonblocking supervisory control problem (BTNSCP) of timed DES, and characterize its solvability in terms of a new concept of {\em bounded-time language completability}.
Then we present an approach to compute the maximally permissive 
solution to the new BTNSCP.
\end{abstract}


\section{Introduction}

In standard supervisory control of (timed) discrete-event systems (DES) (\cite{RamWon87,WonRam87,RamWon89,Wonham16a,CaiWon20,WonCaiRud18,CasLaf08, BraWon94}), and other extensions and applications on nonblocking supervisory control (e.g. \cite{KumarShayman1994, FabianKumar:1997, MaWonham:2005, MalikLudec:2008,BalemiEt:1993, BrandinCharbonnier:1994, Malik:2003}), the plant to be controlled is  modelled by finite-state automata and {\em marker states} are used to represent `desired states'. A desired state can be a goal location, a start/home configuration, or a task completion (\cite{Elienberg:1974, Wonham16a}).  Besides enforcing all imposed control specifications, a {\em nonblocking supervisor} ensures that every system trajectory can reach a marker state (in a finite number of steps). As a result, the system under supervision may always be able to reach a goal, return home, or complete a task.

The nonblocking property only {\em qualitatively} guarantees finite reachability of marker states. There is no given bound on the number of steps or time for reaching marker states, so it can take an arbitrarily large (though finite) number of steps or time before a marker state is reached. Consequently, this qualitatively nonblocking property might not be sufficient for many practical purposes, especially when there are prescribed bounds (of transition steps or time) for reaching desired states. To address this issue, recently \cite{ZhangWangCai:2021} and \cite{ZhangCaiAutomatica:2024} proposed a {\em quantitatively} nonblocking property to capture the practical requirement that marker states be reached within a prescribed number of (transition) steps.
By this property, in the worst case, each reachable state can reach a marker state in no more than a prescribed number of steps.
However, in many practical cases, tasks (represented by marker states) may often be required to be completed within a specific bounded {\it time}.
For example, a rescue vehicle is required not only to reach a goal location but to do so within a given time; a warehouse AGV is expected not only to return to a self-charging area but to do so periodically with a predetermined period; and a production machine is required not only to complete a task (e.g. processing a workpiece) but also to do so within a prescribed time. In Section II below, we will present a more detailed motivation example.

With the above motivation, we extend the concept of quantitative nonblockingness from untimed DES (the elapse of
time is not explicitly modelled) to the framework of timed DES (TDES) (\cite{Wonham16a, BraWon94}), where the occurrence of a time unit is described by the occurrence of a $tick$ event. Thus the bound of time for reaching a subset of marker states can be represented by the number of $ticks$. In this framework, we measure the `maximal time' between the reachable states and the specific subset of marker states, and this is done by counting the number of $ticks$ (rather than the number of all events) in every string leading a reachable state to one of the marker states in the specified subset.  More specifically, consider
a TDES plant modelled by a $tick$-automaton (which is a finite-state automaton with a special $tick$ event), a cover $\{Q_{m,i}|i \in \mathcal{I}\}$ ($\mathcal{I}$ is an index set) on the set of marker states of the $tick$-automaton, and
let $N_i$ be a finite positive integer which denotes the required number of $ticks$ to reach marker states in $Q_{m,i}$. We define a \emph{bounded-time nonblocking property} (with respect to $\{(Q_{m,i}, N_i)|i \in \mathcal{I}\}$) of the $tick$-automaton if from every reachable state of this $tick$-automaton, the number of $ticks$ included in each string that leads the reachable state to marker states in $Q_{m,i}$ is smaller or equal to $N_i$. That is, in the worst case,
for every marker state subset $Q_{m,i}$, every reachable state can reach $Q_{m,i}$ in no more than $N_i$ $ticks$ following any string.

Moreover, we formulate a new \emph{bounded-time nonblocking supervisory control problem of TDES} by requiring a supervisory control solution to be implementable by a bounded-time nonblocking $tick$-automaton. To solve this problem, we present a necessary and sufficient condition by identifying a language property called \emph{bounded-time completability}. The latter roughly means that in the worst case, for every
sublanguage $K_i$ ($i \in \mathcal{I}$) (defined according to a particular type of task corresponding to $Q_{m,i}$) of a given language $K$, every string in the closure of $K$ can be extended to a string in the sublanguage $K_i$ in no more than $N_i$ $ticks$. Further we show that this bounded-time language completability is closed under arbitrary set unions, and together with language controllability which is also closed under unions, a maximally permissive solution exists for the newly formulated bounded-time nonblocking supervisory control problem of TDES. Finally we design effective algorithms for the computation of such an optimal solution.

Detailed literature review on work related to quantitative nonblockingness of untimed DES is referred in \cite{ZhangWangCai:2021, ZhangCaiAutomatica:2024}.
For TDES, time-bounded liveness is similar to our concept of bounded-time nonblockingness;  time-bounded liveness is introduced in \cite{BerardEt:2001} to express and verify constraints on the delays described by exact time, and there are algorithms (e.g. \cite{BonakdarpourKulkarni:2006})  and tools (e.g. KRONORS \cite{Kronos:2002}) that can verify the time-bounded liveness. However, unlike our supervisory control problems of TDES, uncontrollable events and maximal permissiveness of control strategies are not considered.

The contributions of this paper are as follows.
\begin{itemize}
\item First, for a given set of tasks each represented by a subset $Q_{m,i}$ of marker states and associated with a positive integer $N_i$, we propose a new property of $tick$-automaton, called bounded-time nonblockingness. This property quantifies the standard nonblocking property by capturing the practical requirement that each task be completed within the bounded $N_i$ $ticks$, which extends the concept of quantitative nonblockingness in untimed DES by counting the occurrences of the event $tick$.
\item Second, we formulate a new bounded-time nonblocking supervisory control problem of TDES, and characterize its solvability by a bounded-time language completability in addition to language controllability. This problem and its solvability condition are again generalizations of the standard supervisory control problem and solvability condition, and extensions of the quantitatively nonblocking supervisory control problem.
\item Third, we prove that the language property of bounded-time completability is closed under arbitrary set unions, and develop a automaton-based algorithm to compute the supremal bounded-time completable sublanguage. Comparing with the algorithm in \cite{ZhangCaiAutomatica:2024} for computing the supremal quantitatively completable sublanguage, the range of the counter $d$ and the rules for updating $d$ are different; the details are explained in Section 4.
\item Fourth, we present a fixpoint algorithm to compute the supremal controllable and bounded-time completable sublanguage of
a given (specification) language, which synthesizes an optimal (maximally permissive) supervisory control solution for the bounded-time nonblocking supervisory control problem of TDES.
\end{itemize}

The rest of this paper is organized as follows. Section 2 reviews the
nonblocking supervisory control theory of TDES and presents a motivating example for this work.
Section 3 introduces the new concept of bounded-time nonblocking $tick$-automaton, and formulates the bounded-time nonblocking supervisory control
problem (BTNSCP) of TDES. Section 4 presents a necessary and sufficient condition for the solvability of BTNSCP in terms of a new concept of bounded-time language completability, and develops an algorithm to compute the supremal bounded-time completable sublanguage. Section 5 presents a solution to the bounded-time nonblocking supervisory control
problem. Finally Section 6 states our conclusions.

\section{Preliminaries and Motivating Example}

In this section, we review the standard nonblocking supervisory
control theory of TDES in the Brandin-Wonham framework (\cite{BraWon94};\cite[Chapter~9]{Wonham16a}), and present a motivating
example for our work.

\subsection{Nonblocking Supervisory Control of TDES}
First consider the untimed DES model ${\bf G}_{act} = (A, \Sigma_{act}, \delta_{act}, a_0, A_m)$;
here $A$ is the finite set of {\it activities}, $\Sigma_{act}$
the finite set of {\it events}, $\delta_{act}:A \times \Sigma_{act}
\to A$ the (partial) {\it transition function}, $a_0 \in
A$ the {\it initial activity}, and $A_m \subseteq A$ the set
of {\it marker activities}. Let $\mathbb{N}$ denote the set of
natural numbers $\{0,1,2,...\}$, and introduce \emph{time} into ${\bf
G}_{act}$ by assigning to each event $\sigma \in \Sigma_{act}$ a
{\it lower bound} $l_{{\bf G},\sigma} \in \mathbb{N}$ and an {\it upper
bound} $u_{{\bf G},\sigma} \in \mathbb{N} \cup\{\infty\}$, such that
$l_{{\bf G},\sigma} \leq u_{{\bf G},\sigma}$.
Also introduce a distinguished event, written $tick$, to
represent ``tick of the global clock''. Then the TDES model is described by a $tick$-automaton
\begin{equation} \label{eq:G}
{\bf G} := (Q, \Sigma, \delta, q_0, Q_m),
\end{equation}
which is constructed from ${\bf G}_{act}$ (the detailed construction rules are referred to \cite{BraWon94} and
\cite[Chapter~9]{Wonham16a})
such that $Q$ is the finite set of
\emph{states}, $\Sigma := \Sigma_{act} \dot\cup \{tick\}$
the finite set of events, $\delta:Q\times\Sigma \rightarrow Q$
the (partial) {\it state transition function}, $q_0$ the {\it
initial state}, and $Q_m$ the set of {\it marker states}.

Let $\Sigma^*$ be the set of all finite strings of elements in
$\Sigma = \Sigma_{act} \dot\cup \{tick\}$, including the empty
string $\epsilon$.  The transition function $\delta$ is extended to
$\delta:Q\times \Sigma^* \rightarrow Q$ in the usual way. The {\it
closed behavior} of $\bf G$ is the language
$L({\bf G}) := \{s \in \Sigma^*|\delta(q_0,s)!\}$
and the {\it marked behavior} is
$L_m({\bf G}) := \{s \in L({\bf G})| \delta(q_0, s) \in Q_m\} \subseteq L({\bf G})$.
Let $K \subseteq \Sigma^*$ be a language; its {\it prefix
closure} is $\overline{K} := \{s\in\Sigma^*|
(\exists t\in\Sigma^*)~st \in K\}$. $K$ is said to be $L_m({\bf G})$-{\it closed} if
$\overline{K} \cap L_m({\bf G}) = K$.

For $tick$-automaton $\bf G$ as in (\ref{eq:G}), a state $q \in Q$ is \emph{reachable} if there is a string $s \in L({\bf G})$ such that $q = \delta(q_0, s)$; state $q \in Q$ is \emph{coreachable}
 if there is a string $s \in \Sigma^*$ such that $\delta(q, s)!$ and $\delta(q, s) \in Q_m$.
We say that ${\bf G}$ is \emph{nonblocking} if every reachable state in ${\bf G}$ is coreachable.
In fact ${\bf G}$ is \emph{nonblocking} if and only if
$\overline{L_m({\bf G})} = L({\bf G})$.

To use $\bf G$ in (\ref{eq:G}) for supervisory control, first
designate a subset of events, denoted by $\Sigma_{hib} \subseteq \Sigma_{act}$,
to be the {\it prohibitible} events which can be disabled by an external supervisor.
Next, and specific to TDES, specify a subset of {\it forcible} events,
denoted by $\Sigma_{for} \subseteq \Sigma_{act}$, which can {\it preempt}
the occurrence of event $tick$.

Now it is convenient to define the {\it controllable} event set $\Sigma_c :=
\Sigma_{hib}~\dot\cup~\{tick\}$.
The {\it uncontrollable} event set is $\Sigma_{uc} := \Sigma \setminus \Sigma_c$.
A {\it supervisory control} for $\bf G$ is any map $V:L({\bf G})\rightarrow \Gamma$,
where $\Gamma := \{\gamma \subseteq \Sigma \mid \gamma \supseteq \Sigma_{uc}\}$. Then the
{\em closed-loop system} is denoted by $V/{\bf G}$, with closed behavior $L(V/{\bf G})$ defined as follows:
\begin{align*}
\mbox{(i) } &\epsilon \in L(V/{\bf G}); \\
\mbox{(ii) } &s \in L(V/{\bf G}) \ \&\ \sigma \in V(s) \ \&\ s\sigma \in L({\bf G}) \Rightarrow s\sigma \in L(V/{\bf G});\\
\mbox{(iii) }& \text{no other strings belong to } L(V/{\bf G}).
\end{align*}
On the other hand, for any sublanguage $K \subseteq L_m({\bf G})$, the closed-loop system's marked behavior $L_m(V/{\bf G})$ is given by\footnote{With this definition of $L_m(V/{\bf G})$, the supervisory control $V$ is also known as a {\em marking supervisory control} for $(K, {\bf G})$ (\cite{Wonham16a}).}
\begin{align*}
L_m(V/{\bf G}) := K \cap L(V/{\bf G}).
\end{align*}
The closed behavior $L(V/{\bf G})$ represents the strings generated by the plant ${\bf G}$ under the control of $V$,
while the marked behavior $L_m(V/{\bf G})$ represents the strings that have some special significance, for instance
representing `task completion'. We say that $V$ is {\it nonblocking} if
\begin{align}
\overline{L_m(V/{\bf G})} = L(V/{\bf G}).
\end{align}

A sublanguage $K \subseteq L_m({\bf G})$ is {\it controllable} if, for
all $s \in \overline{K}$,
\begin{eqnarray*} 
Elig_K(s)\supseteq
\left\{
   \begin{array}{lcl}
      Elig_{\bf G}(s)\cap(\Sigma_{uc} \dot{\cup}\{tick\}) &\text{if} &~~Elig_K(s)\cap \Sigma_{for} = \emptyset,\\
      ~~~~~~~~~~~~~~~~~\\
      Elig_{\bf G}(s)\cap\Sigma_{uc}  &\text{if} &~~Elig_K(s)\cap \Sigma_{for} \neq \emptyset
      ~~~~~~~~~~~~~~~~~
   \end{array}
\right.
\end{eqnarray*}
where $Elig_{\bf G}(s):= \{\sigma \in \Sigma|s\sigma \in L({\bf G})\}$
and $Elig_K(s):= \{\sigma \in \Sigma|s\sigma \in \overline{K}\}$
are the subset of eligible events after string $s$ in $L({\bf G})$ and $K$ respectively.

The following is a central result of nonblocking supervisory control theory (\cite{BraWon94,Wonham16a}).

\begin{thm} \label{thm:sct}
Let $K\subseteq L_m({\bf G})$, $K \neq \emptyset$. There exists a nonblocking (marking) TDES supervisory control $V$ (for $(K, {\bf G})$) such that $L_m(V/{\bf G}) = K$ if and only if $K$ is controllable. Moreover, if such a nonblocking TDES supervisory control $V$ exists, then it may be implemented by a nonblocking supervisor 
${\bf SUP}$, i.e. $L_m({\bf SUP}) = L_m(V/{\bf G})$. \hfill $\diamond$
\end{thm}

Further, the property of language controllability is closed
under set union. Hence for any language $K \subseteq L_m({\bf G})$ (whether or not controllable), the set $\mathcal{C}(K) = \{K' \subseteq K \mid K' ~\text{is controllable wrt.} ~{\bf G} \}$ contains a unique supremal element denoted by $\sup \mathcal{C}(K)$.  Whenever $\sup \mathcal{C}(K)$ is nonempty, by Theorem~\ref{thm:sct} there exists a nonblocking supervisory control $V_{\sup}$ that satisfies $L_m(V_{\sup}/{\bf G}) = \sup \mathcal{C}(K)$ and may be implemented by a nonblocking supervisor ${\bf SUP}$ with \[L_m({\bf SUP}) = L_m(V_{\sup}/{\bf G}).\]

\subsection{Motivating Example}

Nonblockingness of supervisory control $V$ describes a general requirement that every string
generated by the closed-loop system $V/{\bf G}$ can be completed to
a marked string in indefinite (finite but unbounded) number of transitions, which may include indefinite number of $ticks$.
Namely, the time needed for completing a task is unbounded. However, in many real-world applications, it is often required that a task be completed in a prescribed, bounded time. As an illustration, we present the following example (adapted from Example 1 in \cite{ZhangCaiAutomatica:2024}).

\begin{figure}[!t]
\centering
    \includegraphics[scale = 0.6]{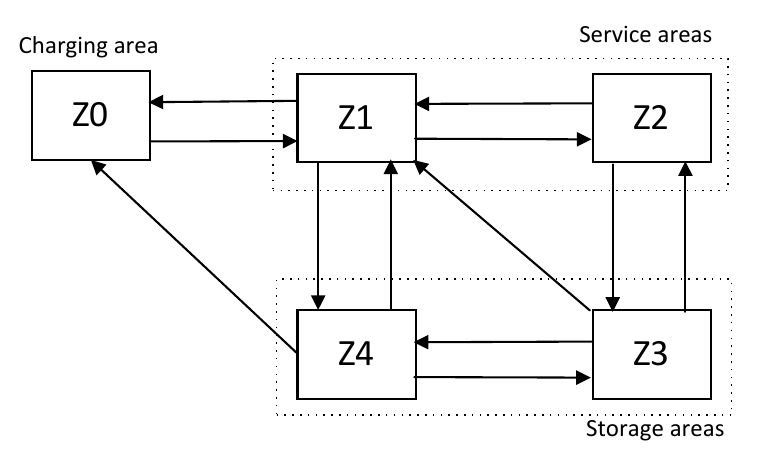}
\caption{Routes of the vehicle} \label{fig:Routes}
\end{figure}

\begin{example}
Consider an autonomous vehicle for package collecting and delivery in a local region.
The vehicle can move in five zones numbered 0--4, following the routes displayed in Fig.~\ref{fig:Routes}.
Zone~0 is the charging area for the vehicle to charging its battery. Zones 1 and 2 are
two service areas for customers where the customers can both receive packages from the
vehicle and call the vehicle to come to collect packages to be sent. Zones 3 and 4 are the storage areas
for incoming and outgoing packages. Namely, the task of the vehicle is to send packages in the storage
areas (zones 3 and 4) to the service areas (zones 1 and 2), and collect packages from the service areas
and store them into the storage areas. Also, the vehicle must be able to make a self-charging when it is running
out of battery.

Each movement of the vehicle from one zone to the next is represented by two timed events: one represents the leaving from one zone and the other represents the arriving to the next.
Both events have lower and upper bounds. As one example, assuming that the lower and upper bounds for the vehicle leaving from zone 1 for zone 0 are 1 ($tick$) and $\infty$ respectively and one time unit (a $tick$) represents 2 minutes, the vehicle may start to move from zone 1 at any time after 1 $tick$ elapsed. For another example, assuming that the lower and upper bounds for the vehicle arriving zone 0 from zone 1 are 1 ($tick$) and 2 ($ticks$) respectively, the vehicle may arrive zone 0 in 1 or 2 $ticks$ if it has left zone 1. Similarly, we assign lower and upper times bounds to other timed events, as displayed in Table~1.
\begin{table}[!t]\label{tab:eventlabel} \caption{Event representations of each vehicle route, and the corresponding lower and upper bounds. Notation: $Zi$ ($i = 0,1,2,3,4,5$) represents zone $i$.}
\centering
{
\begin{tabular}{|c|c|c|}
  \hline
   route & event label & (lower bound, upper bound) \\
   \hline
    Leave $Z0$ for $Z1$ & 1 & (1,$\infty$) \\
   \hline
    Arrive $Z1$ from $Z0$ & 2 & (1,1) \\
   \hline
   Leave $Z1$ for $Z0$ & 11 & (0,$\infty$) \\
   \hline
    Arrive $Z0$ from $Z1$ & 12 & (1,1) \\
   \hline
    Leave $Z1$ for $Z2$ & 13 & (0,$\infty$) \\
   \hline
    Arrive $Z2$ from $Z1$ & 14 & (1,2) \\
   \hline
    Leave $Z1$ for $Z4$ & 15 & (0,$\infty$) \\
   \hline
    Arrive $Z4$ from $Z1$ & 16 & (1,1) \\
  \hline
   Leave $Z2$ for $Z1$ & 21 & (0,$\infty$) \\
   \hline
    Arrive $Z1$ from $Z2$ & 22 & (1,2) \\
  \hline
   Leave $Z2$ for $Z3$ & 23 & (0,$\infty$) \\
   \hline
    Arrive $Z3$ from $Z2$ & 24 & (1,1) \\
  \hline
   Leave $Z3$ for $Z1$ & 31 & (0,$\infty$) \\
   \hline
    Arrive $Z1$ from $Z3$ & 32 & (1,2) \\
  \hline
   Leave $Z3$ for $Z2$ & 33 & (0,$\infty$) \\
   \hline
    Arrive $Z2$ from $Z3$ & 34 & (1,1) \\
  \hline
      Leave $Z3$ for $Z4$ & 35 & (0,$\infty$) \\
   \hline
    Arrive $Z4$ from $Z3$ & 36 & (1,2) \\
  \hline
  Leave $Z4$ for $Z0$ & 41 & (0,$\infty$) \\
   \hline
    Arrive $Z0$ from $Z4$ & 42 & (2,2) \\
       \hline
   Leave $Z4$ for $Z1$ & 43 & (0,$\infty$) \\
   \hline
    Arrive $Z1$ from $Z4$ & 44 & (1,1) \\
  \hline
      Leave $Z4$ for $Z3$ & 45 & (0,$\infty$) \\
   \hline
    Arrive $Z3$ from $Z4$ & 46 & (1,1) \\
  \hline
\end{tabular}
}
\end{table}

We model the movement of the autonomous vehicle by an untimed DES model ${\bf G}_{act}$ with transition graph displayed in Fig.~\ref{fig:ATG}. States~0 (charging area Z0), 2 and 5 (service areas Z1 and Z2) are chosen to be the marker states, and the reaching of marker states represents that the vehicle arrives the corresponding areas. The timed DES model ($tick$-automaton) $\bf G$ of the vehicle can be generated according to the construction rules in (\cite{BraWon94,Wonham16a}); its transition graph is displayed as in Fig.~\ref{fig:SUP}. Assume that events 1, 11, 13, 15, 21, 23, 31, 33, 35, 41, 43 and 45 are both prohibitible and forcible.
\begin{figure}[!t]
\centering
    \includegraphics[scale = 0.4]{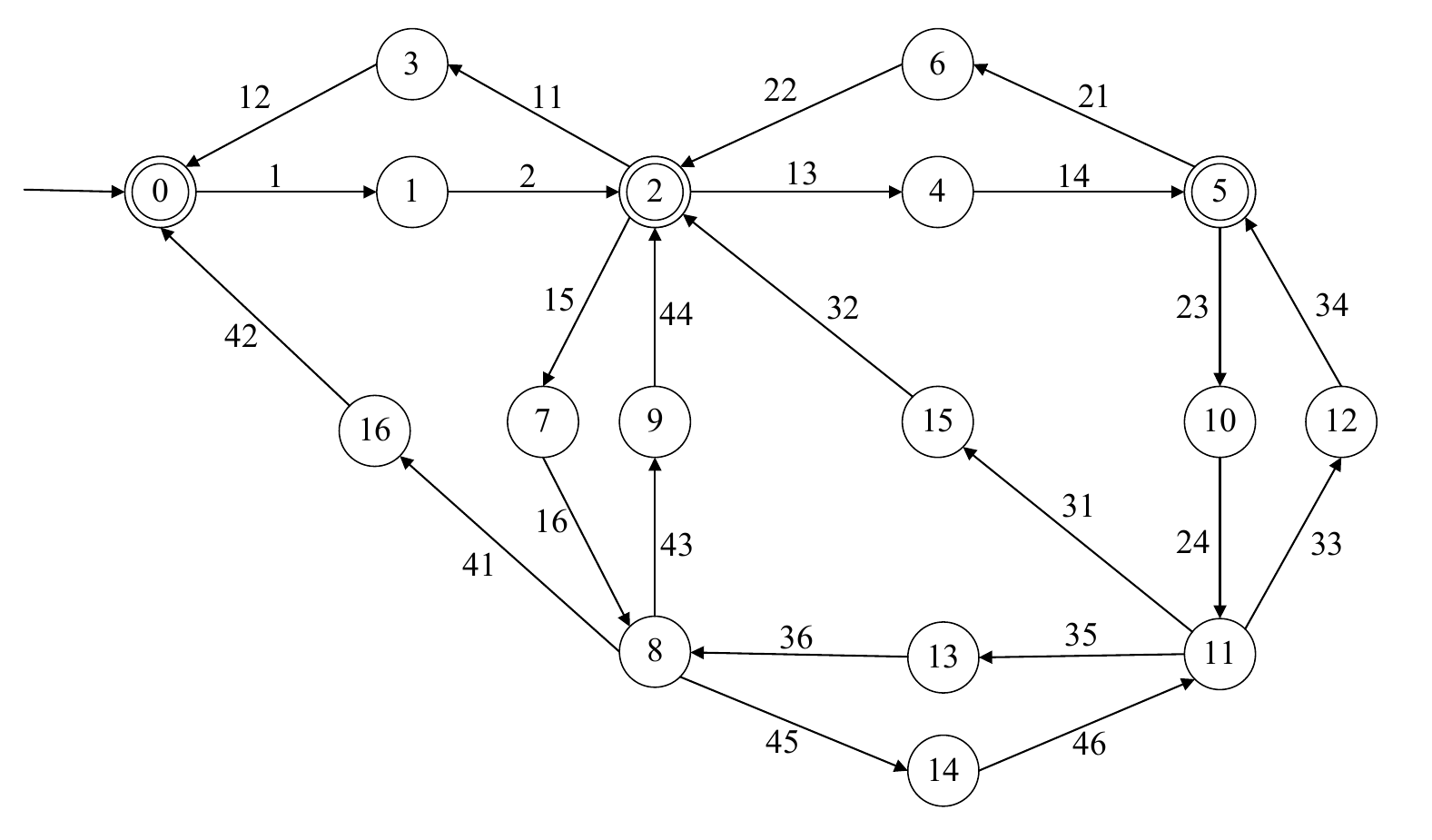}
\caption{Transition graph of ${\bf G}_{act}$. In the transition graph, states 0, 2, 5, 8, and 11 represent that the vehicle stays
at the area Z0, Z1, Z2, Z4 and Z3 respectively; other states represent that the vehicle is in the process of
moving from one zone to the next, e.g. state 1 represents that the vehicle is in the process of moving from zone 0 to zone 1.} \label{fig:ATG}
\end{figure}

\begin{figure}[!t]
\centering
    \includegraphics[scale = 0.38]{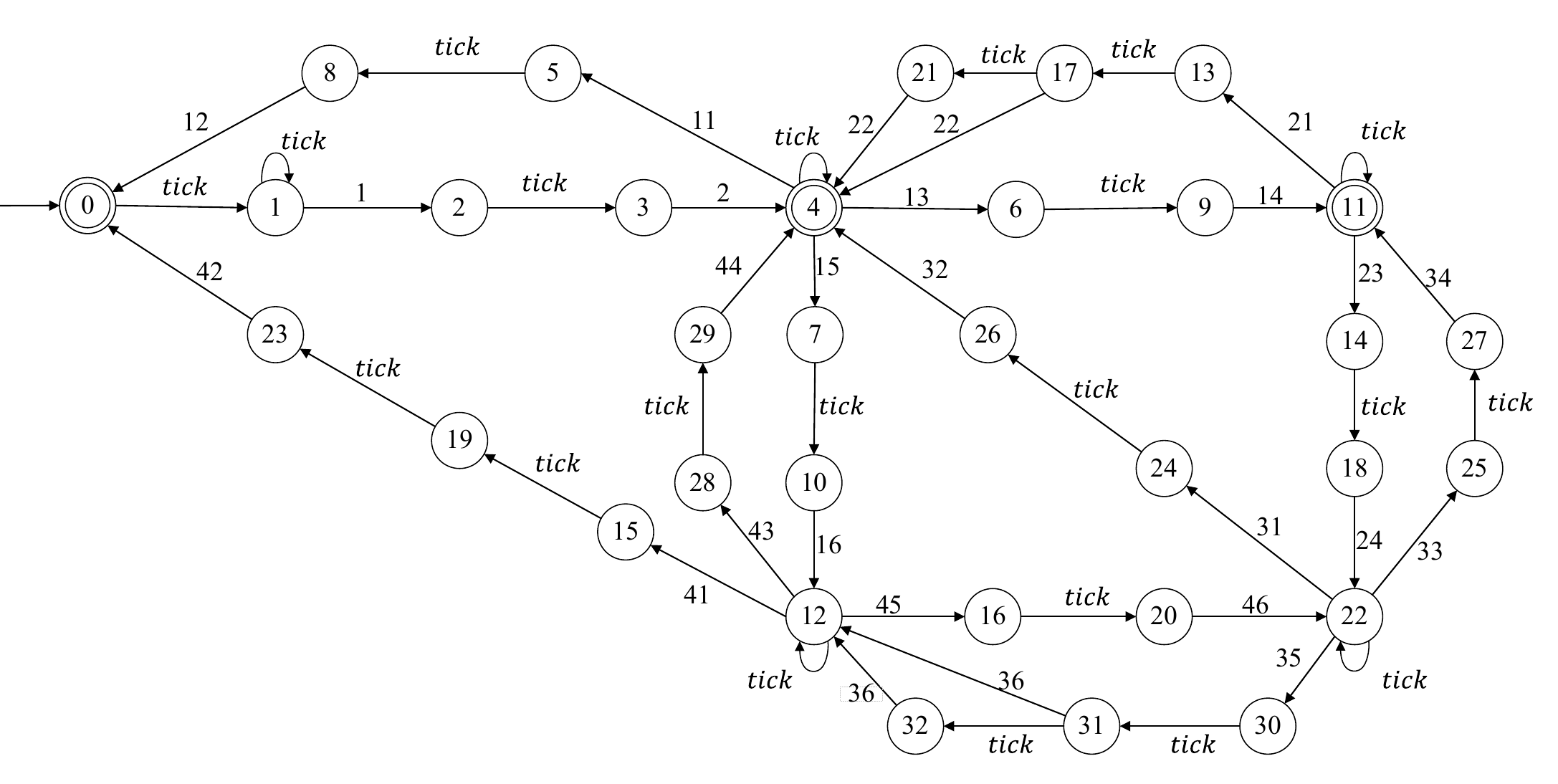}
\caption{Transition graph of ${\bf G}$} \label{fig:TTG}
\end{figure}

Suppose that due to road maintenance, the (directed) route
\\

~~~~~~~~~~~~~~~~~~ zone~3 $\rightarrow$ zone~4 $\rightarrow$ zone~1
\\

\noindent is not usable. This constraint is imposed as a safety specification. We further consider a temporal specification that
after arriving {\color{red}zone 4}, the vehicle should collect or store the package, and leave this area in 1 $tick$.
To satisfy these two specifications, a nonblocking supervisory control can be synthesized (\cite{BraWon94,Wonham16a}), and implemented by a nonblocking supervisor {\bf SUP} as displayed in Fig.~\ref{fig:SUP}.
This  ${\bf SUP}$ needs to (1) disable event 35 at state~22 (zone 3), and event 43 at states~12 and 28 (zone 4) (to satisfy the safety specification);
and (2) preempt event $tick$ at state 28 by events 41 or 45 (to satisfy the temporal specification).
Moreover, since ${\bf SUP}$ is nonblocking, every reachable state can reach the marker states~0, 4, 11 (representing zones 0, 1 and 2 respectively) in finite number of $ticks$.

\begin{figure}[!t]
\centering
    \includegraphics[scale = 0.38]{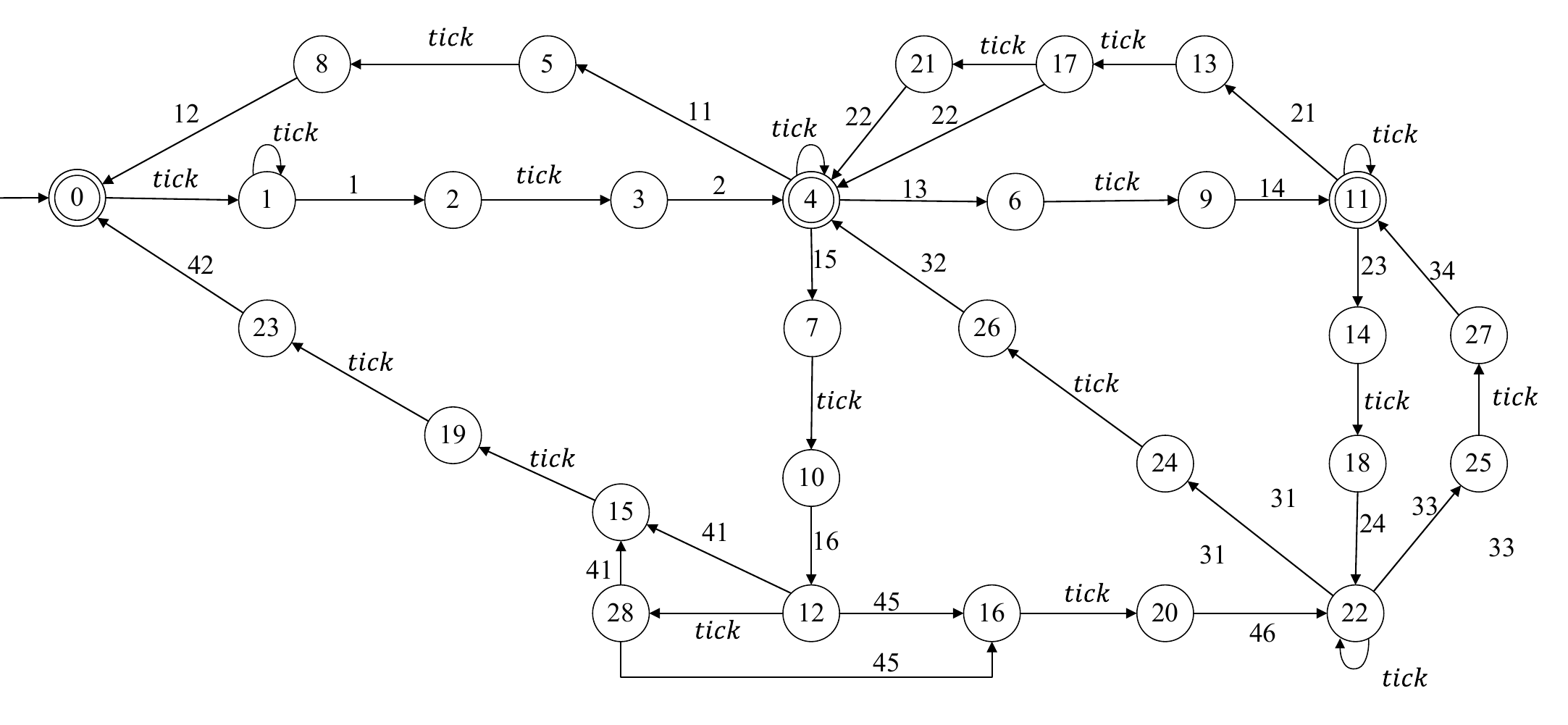}
\caption{Transition graph of $\bf SUP$} \label{fig:SUP}
\end{figure}

Now consider two additional requirements:

\begin{enumerate}[{\rm (i)}]
 \item Every package sent to customers  must be delivered by the vehicle to either one of the two service areas (zone~1 or 2) within
 10 minutes (5 $ticks$); and whenever a customer calls for package collection, the vehicle must reach zone~1 or 2 within 10 minutes;

 \item {The vehicle must be able to return to zone 0 for charging its battery within 18 minutes (9 $ticks$).}
\end{enumerate}

Note that the above requirements are different from the temporal specification: the latter imposes temporal constraints on event occurrences;
but the above requirements impose bounded time constraints on arriving marker states.

The nonblocking supervisor $\bf SUP$ in Fig.~\ref{fig:SUP} fails to fulfill the above requirements,
because if the vehicle is at zone~3 (state 11), it is not guaranteed that the vehicle can move to zone~1 (state 2) in 5 $ticks$ (it may stay at zone~3 in any number of $ticks$).

Hence, we need a new method that can count the exact time (number of $ticks$) needed for completing each task,
and design a supervisor to satisfy the bounded-time requirement. To address this issue, we adopt an idea similar to that in \cite{ZhangCaiAutomatica:2024}
which synthesizes quantitatively nonblocking supervisors where the transitions caused by all events are counted,
but with a novel change that the new algorithm must distinguish $tick$ and non-$tick$ events and count only $tick$.

\end{example}

\section{Bounded-Time Nonblocking Supervisory Control Problem Formulation}


We start by introducing a new concept that {\em quantifies} the nonblocking property of a $tick$-automaton.

Let ${\bf G} = (Q,\Sigma,\delta,q_0,Q_m)$ be a $tick$-automaton (modelling the TDES plant to be controlled) as in (\ref{eq:G}) and assume that ${\bf G}$ is nonblocking (i.e. every reachable state of ${\bf G}$ is also coreachable).
Bring in a cover  $\mathcal{Q}_{\bf G}$ on the marker state set $Q_m$ as follows:
\begin{align} \label{eq:part}
\mathcal{Q}_{\bf G} := \{Q_{m,i} \subseteq Q_m | i \in \mathcal{I}\}.
\end{align}
Here $\mathcal{I}$ is an index set, $Q_{m,i} \neq \emptyset$ for each $i \in \mathcal{I}$, 
and $\bigcup \{Q_{m,i} | i \in \mathcal{I}\} $ $=  Q_m$.
This cover $\mathcal{Q}_{\bf G}$ represents a classification of different types of marker states. For example, the three marker states 0, 4, 11 in Example~2.2 can be classified into two types: $Q_{m,1} = \{4,11\}$ meaning completion of a package collecting/delivery task, whereas $Q_{m,2} = \{0\}$ meaning battery charging.

Fix $i \in \mathcal{I}$ and let  $q \in Q \setminus Q_{m,i}$ be an arbitrary state in $Q$ but not in $Q_{m,i}$.
We define the set of all strings that lead $q$ to $Q_{m,i}$ for the first time, namely
\begin{align*}
C(q,Q_{m,i}):=\{s \in \Sigma^*| \delta(q,s)! ~\&~\delta(q,s) \in Q_{m,i}~\&
     (\forall s' \in \overline{s}\setminus\{s\}) ~\delta(x,s') \notin Q_{m,i}\}.
\end{align*}
If $q \in Q_{m,i}$, we define $C(q,Q_{m,i}) := \{\epsilon\}$.

Now associate $Q_{m,i}$ with a finite positive integer $N_i$, and consider an arbitrary state in $q \in Q$.
Denote by $\#s(tick)$ the number of $tick$ occurred in string $s$.
We say that state $q$ is {\em $N_i$-$tick$ coreachable} (wrt. $Q_{m,i}$) if
\begin{align*}
{\rm (i)} ~~ & C(q, Q_{m,i}) \neq \emptyset; \mbox{ and}\\
{\rm (ii)} ~~& (\forall s \in C(q,Q_{m,i})) ~\#s(tick) \leq N_i.
\end{align*}
Condition (i) requires that there exists at least one string $s \in \Sigma^*$ leading $q$ to a marker state in $Q_{m,i}$.
Condition (ii) means that all strings that lead $q$ to $Q_{m,i}$ for the first time include at most $N_i$ $ticks$.
Intuitively, condition~(ii) means that in the worst case, it takes $N_i$ $ticks$ from state $q$ to arrive a marker state in $Q_{m,i}$.
Hence if $Q_{m,i}$ represents a task completion, then condition~(ii) means that in the worst case, it takes $N$ $ticks$ from state $q$ to complete the task.

\begin{remark}
We remark here that a $tick$-automaton $\bf G$ is \emph{activity-loop-free} (\cite{Wonham16a}) (i.e $(\forall q \in Q)(\forall s\in\Sigma_{act}^*\setminus\{\epsilon\})\delta(q,s)\neq q$). Thus all loops (the strings visiting a state repeatedly) in $\bf G$ include at least one $tick$.
It is easily verified that all the $tick$-automata representing the sublanguages of $L_m({\bf G})$ are also activity-loop-free,
and thus the $tick$-automata mentioned in this paper are activity-loop free.

\hfill $\diamond$

\end{remark}

Now we introduce the new concept of bounded-time nonblockingness of a $tick$-automaton.
\smallskip
\begin{defn} \label{defn:snnbg}
Let ${\bf G} = (Q, \Sigma, \delta, q_0, Q_m)$ be a $tick$-automaton, {$\mathcal{Q}_{\bf G} = \{Q_{m,i} | i \in \mathcal{I}\}$ a cover on $Q_m$ as defined in (\ref{eq:part}), and $N_i$ a positive integer associated with each $Q_{m,i} \in \mathcal{Q}_{\bf G}$.}
We say that ${\bf G}$ is \emph{bounded-time nonblocking} wrt. $\{(Q_{m,i}, N_i)|i \in \mathcal{I}\}$ if for every $i \in \mathcal{I}$
and every reachable state $q \in Q$, $q$ is $N_i$-$tick$ coreachable (wrt. $Q_{m,i}$).
\end{defn}
\smallskip

In words, a bounded-time nonblocking $tick$-automaton requires that every state $q$ can reach every subset $Q_{m,i}$ of marker states within $N_i$ $ticks$. Compared with quantitatively nonblockingness
of untimed automaton in \cite{ZhangCaiAutomatica:2024}, bounded-time nonblockingness of $tick$-automaton is focused on
the time for reaching marker states being bounded, rather than in \cite{ZhangCaiAutomatica:2024} that the total number of transitions (caused by any events) for reaching marker state are bounded.

Next we define the bounded-time nonblocking property of a supervisory control $V$ for TDES. For this, we first introduce a new concept called {\em bounded-time completability}.

Let $K \subseteq L_m({\bf G})$ be a sublanguage of $L_m({\bf G})$.
{For each marker state subset $Q_{m,i} \in \mathcal{Q}_{\bf G}$ define
\[L_{m,i}({\bf G}) := \{s \in L_m({\bf G}) | \delta(q_0, s) \in Q_{m,i}\}\]
i.e. $L_{m,i}({\bf G})$ represents the marked behavior of ${\bf G}$ wrt. $Q_{m,i}$. Then for each $i \in \mathcal{I}$, define \[K_i := K \cap L_{m,i}({\bf G}).\]

For an arbitrary string $s \in \overline{K} \setminus K_i$, define the set of strings that lead $s$ to $K_i$ for the first time:
\begin{align} \label{eq:PKL}
M_{K,i}(s):=\{t \in \Sigma^* \mid st \in K_i
(\forall t' \in \overline{t} \setminus\{t\}) st' \notin K_i \}.
\end{align}
If already $s \in K_i$, we define $M_{K,i}(s) :=\{\epsilon\}$.

\smallskip
\begin{defn} \label{defn:sncomp}
Let ${\bf G} = (Q, \Sigma, \delta, q_0, Q_m)$ be a $tick$-automaton, $K  \subseteq L_m({\bf G})$ a sublanguage, {$\mathcal{Q}_{\bf G} = \{Q_{m,i} | i \in \mathcal{I}\}$ a cover on $Q_m$ as defined in (\ref{eq:part}), and $N_i$ a positive integer associated with each $Q_{m,i} \in \mathcal{Q}_{\bf G}$.}
For a fixed $i \in \mathcal{I}$, we say that $K$ is \emph{bounded-time completable} wrt. $(Q_{m,i}, N_i)$ if
for all $s \in \overline{K}$,
\begin{align*}
{\rm (i)}&~ M_{K,i}(s) \neq \emptyset;\\
{\rm (ii)}&~ (\forall t \in M_{K,i}(s))~ \#t(tick) \leq N_i.
\end{align*}
Moreover if $K$ is bounded-time completable wrt. $(Q_{m,i}, N_i)$ for all $i \in \mathcal{I}$, we say that $K$ is {\em time-quuantitatively completable} wrt. $\{(Q_{m,i}, N_i)|i \in \mathcal{I}\}$.
\end{defn}
\smallskip

If $K$ is bounded-time completable  wrt. $\{(Q_{m,i}, N_i)|i \in \mathcal{I}\}$, then for every $i \in \mathcal{I}$, every
string $s \in \overline{K} $ may be extended to a string in $K_i (= K \cap L_{m,i}({\bf G}))$  by strings including at most $N_i$ $ticks$. Compared
with quantitative completability in \cite{ZhangCaiAutomatica:2024}, the second condition is different: here
all the strings in $M_{K,i}(s)$ are required to include at most $N_i$ $ticks$, rather
than have length no more than $N_i$ in defining quantitative completability.

The following result characterizes the relation between bounded-time completability of a language and bounded-time nonblockingness of a $tick$-automaton.

\begin{pro} \label{prop:relation}
Let ${\bf G} = (Q, \Sigma, \delta, q_0, Q_m)$ be a $tick$-automaton, $K \subseteq L_m({\bf G})$ a sublanguage, $\mathcal{Q}_{\bf G} = \{Q_{m,i} | i \in \mathcal{I}\}$ a cover on $Q_m$, and $N_i$ a positive integer associated with each $Q_{m,i} \in \mathcal{Q}_{\bf G}$.

{\rm (i)} If $K = L_m({\bf G})$ and ${\bf G}$ is bounded-time nonblocking wrt. $\{(Q_{m,i}, N_i)|i \in \mathcal{I}\}$, then $K$ is
time-quantitatively completable wrt. $\{(Q_{m,i}, N_i)|i \in \mathcal{I}\}$.

{\rm (ii)} If $K \subseteq L_m({\bf G})$ is bounded-time completable wrt. $\{(Q_{m,i}, N_i)|i \in \mathcal{I}\}$, then there exists a $tick$ automaton ${\bf K} = (X, \Sigma, \xi, x_0, X_m)$
such that $L_m({\bf K})=K$ and ${\bf K}$ is bounded-time nonblocking wrt. $\{(X_{m,i}, N_i)|i \in \mathcal{I}\}$,
where $X_{m,i} = \{x_m \in X_m| (\exists s \in \Sigma^*) \xi(x_0, s) = x_m \ \& \  \delta(q_0, s) \in Q_{m,i}\}$.
\end{pro}

The proof of Proposition~\ref{prop:relation} is similar to that of Proposition 4 in \cite{ZhangCaiAutomatica:2024}.
According to Proposition~\ref{prop:relation}, for an arbitrary sublanguage $K \subseteq L_m({\bf G})$ that is bounded-time completable wrt. $\{(Q_{m,i}, N_i)|i \in \mathcal{I}\}$,
we may construct a bounded-time nonblocking (wrt. $\{(X_{m,i}, N_i)|i \in \mathcal{I}\}$) $tick$-automaton ${\bf K}$ representing $K$, i.e. $L_m({\bf K}) = K$.

With the above bounded-time completability of a language, we introduce the bounded-time nonblocking property of a supervisory control for TDES.
\smallskip
\begin{defn} \label{defn:snnbs}
Let ${\bf G} = (Q, \Sigma, \delta, q_0, Q_m)$ be a $tick$-automaton, $K  \subseteq L_m({\bf G})$ a sublanguage, {$\mathcal{Q}_{\bf G} = \{Q_{m,i} | i \in \mathcal{I}\}$ a cover on $Q_m$ as defined in (\ref{eq:part}), $N_i$ a positive integer associated with each $Q_{m,i} \in \mathcal{Q}_{\bf G}$,}
and $V : L({\bf G}) \rightarrow \Gamma$ a (marking) TDES supervisory control (wrt. $(K,{\bf G})$).
We say that $V$ is \emph{bounded-time nonblocking} wrt. {$\{(Q_{m,i}, N_i)|i \in \mathcal{I}\}$} if
\begin{align*}
{\rm (i)} ~~ &V \text{ is nonblocking}; \mbox{ and}\\
{\rm (ii)} ~~&L_m(V/{\bf G}) ~\text{is bounded-time completable wrt. {$\{(Q_{m,i}, N_i)|i \in \mathcal{I}\}$}.}
\end{align*}
\end{defn}
\smallskip

In words, quantitative nonblockingness of a TDES supervisory control $V$ requires not only $V$ being nonblocking (in the standard sense), but also the marked behavior $L_m(V/{\bf G})$ of
the closed-loop system $V/{\bf G}$ being bounded-time completable. According to Proposition~\ref{prop:relation},
$L_m(V/{\bf G})$ can be represented by a bounded-time nonblocking $tick$-automaton.

We are ready to formulate the \emph{Bounded-Time Nonblocking Supervisory Control Problem} of TDES ({\em BTNSCP}):

{\it Consider a TDES plant modelled by a $tick$-automaton ${\bf G} = (Q,\Sigma_c \dot\cup \Sigma_{uc},\delta,q_0,Q_m)$, a specification language $E \subseteq \Sigma^*$, and let  $K := E\cap L_m({\bf G})$,
{$\mathcal{Q}_{\bf G} = $ $ \{Q_{m,i} \subseteq Q_m |i \in \mathcal{I}\}$ a cover on $Q_m$, and $N_i$ a positive integer associated with each $Q_{m,i} \in \mathcal{Q}_{\bf G}$.}
Construct a (marking) TDES supervisory control $V: L({\bf G})\rightarrow \Gamma$ (for ($K,{\bf G}$))
satisfying the following properties:
\begin{itemize}
\item[$\bullet$]  {\bf Safety}. Marked behavior of the closed-loop system $V/{\bf G}$ satisfies the imposed specification $E$ in the sense that {$L_m(V/{\bf G}) \subseteq E\cap L_m({\bf G})$}.

\item[$\bullet$] {\bf Bounded-time nonblockingness}. TDES supervisory control $V$ is bounded-time nonblocking wrt. $\{(Q_{m,i}, N_i)|i \in \mathcal{I}\}$.

\item[$\bullet$] {\bf Maximal permissiveness}. TDES supervisory control $V$ does not restrict more behavior than necessary to satisfy
safety and  bounded-time nonblockingness, i.e. for all
other safe and  bounded-time nonblocking TDES supervisory controls $V'$ it holds that  $L_m(V'/{\bf G}) \subseteq L_m(V/{\bf G})$.
\end{itemize}
}

The BTNSCP is a variation of the the traditional nonblocking supervisory control problem (\cite{BraWon94,Wonham16a}) of TDES, in that the second requirement of bounded-time nonblockingness is stronger than the traditional nonblockingness. Namely, this problem cannot be solved in general by supervisors synthesized using the standard method.
Moreover, the BTNSCP can not be solved by only representing the bounded-time completability by a specification language,
because the synthesis of nonblocking supervisors may remove marker states and thus violate the
bounded-time completability; an example is as follows.

Let ${\bf NG}$ be a generator representing the TDES plant to be controlled, and ${\bf NSPEC}$ the generator representing a specification language imposed
on the plant, as shown in Fig.~\ref{fig:NGSPEC}. Assume that events 13 and 15 are both prohibitable and forcible, i.e. $\Sigma_{hib} = \Sigma_{for} = \{13, 15\}$. It is easily verified that $L_m({\bf NSPEC})$ is controllable, and
thus $\bf NSPEC$ can be acted as a nonblocking supervisor according to the standard nonblocking supervisory control theory of TDES.

\begin{figure}[!t]
\centering
    \includegraphics[scale = 0.4]{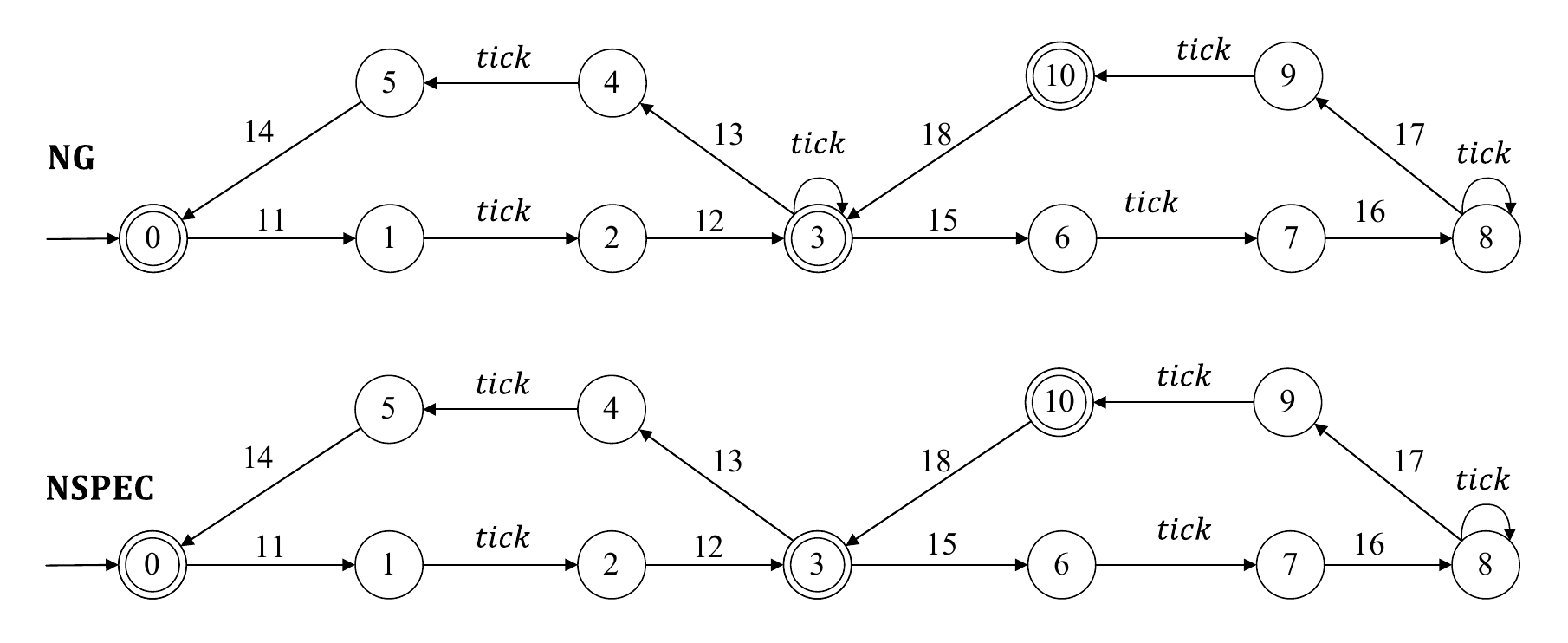}
\caption{Transition graphs of ${\bf NG}$ and ${\bf NSPEC}$} \label{fig:NGSPEC}
\end{figure}
Now consider a cover $\mathcal{Q}_{\bf G} = \{Q_{m,i} \subseteq Q_m |i \in \{1,2\}\}$ with $Q_{m,1} = {1,3}$ and $Q_{m,2} = \{10\}$;
associate with $Q_{m_1}$ and $Q_{m_2}$ two integers $N_1 = 2$ and $N_2 = 4$ respectively.
Let $L_m({\bf NSPEC}')$ be
a language modified from $L_m({\bf NSPEC}')$, which is represented by a generator as shown in Fig.~\ref{fig:NSPEC}. According to the definition of bounded-time completability,
$L_m({\bf NSPEC}')$ is bounded-time completable wrt. $\{(Q_{m,i}, N_i)|i \in \{1,2\}\}$. Namely,
$L_m({\bf NSPEC}')$ replaces $L_m({\bf NSPEC})$ as a new specification having the property of bounded-time completability.

\begin{figure}[!t]
\centering
    \includegraphics[scale = 0.4]{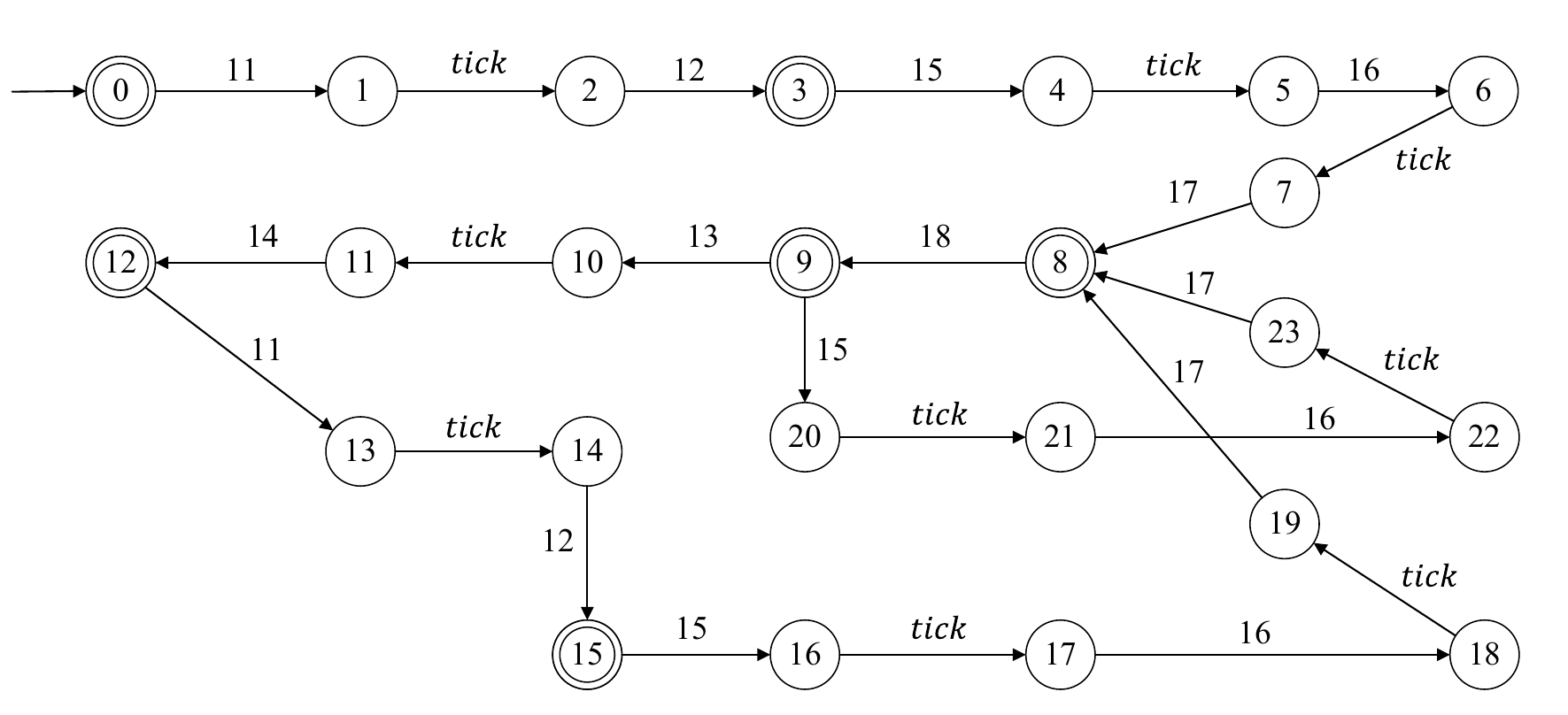}
\caption{Transition graph of ${\bf NSPEC}'$} \label{fig:NSPEC}
\end{figure}

Then, treating $L_m({\bf NSPEC}')$ as a specification imposing on plant $\bf NG$ and applying the standard nonblocking supervisory control theory of TDES, we obtain a nonblocking supervisor ${\bf NSUP}$ as displayed in Fig.~\ref{fig:NSUP} (event $tick$ is removed from state 6 but event 17 is not forcible; then according to
the algorithm of computing supremal controllable sublanguage, we need disable event 15 at state 3 to satisfy the property of nonblockingness).

\begin{figure}[!t]
\centering
    \includegraphics[scale = 0.4]{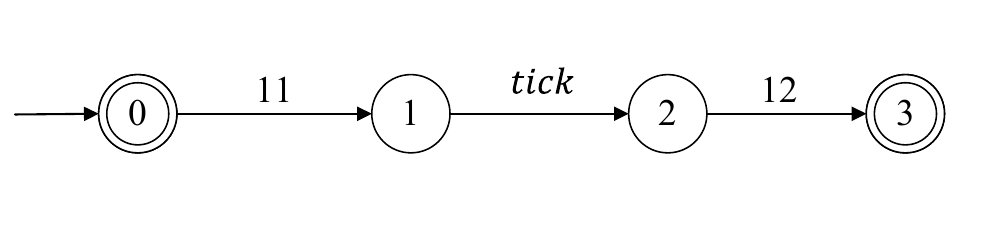}
\caption{Transition graph of ${\bf NSUP}$} \label{fig:NSUP}
\end{figure}

We see that  ${\bf NSUP}$ is not bounded-time completable wrt. $\{(Q_{m,i}, N_i)|i \in \{1,2\}\}$, because starting from state 0 the plant cannot visit marker state 10 in $Q_{m,2}$. The intuition is that
the algorithm of computing the supremal controllable sublanguage removes some transitions from ${\bf NSPEC}'$, which violates the property of bounded-time completability.

However, as we have mentioned in \cite{ZhangCaiAutomatica:2024}, an alternative approach to solve QNSCP is to first compute the supremal quantitatively completable sublanguage
(or represent the property of quantitative completability by a new specification language)
and then compute the supremal controllable and strongly nonblocking sublanguage by the algorithm in \cite{QueirozEt:2005} with the quantitatively completable sublanguage
as input. Similarly, the BTNSCP may be solved by first computing the supremal bounded-time completable sublanguage
(or represent the property of bounded-time completability by a new specification language)
and then computing the supremal controllable and strongly nonblocking sublanguage by extending the algorithm in from untimed DES to TDES.

In \cite{ZhangCaiAutomatica:2024}, we have proposed an effective approach to compute the supremal quantitatively completable and controllable sublanguage,
need not the algorithm of synthesizing strongly nonblocking supervisors.
Hence, in subsequent sections, we will develop solution algorithms based on those in \cite{ZhangCaiAutomatica:2024} to satisfy the new requirement of bounded-time nonblockingness and obtain
the maximal solution to the BTNSCP, need not the extension of strong nonblockingness from untimed
DES to TDES.

\section{Supremal Bounded-Time Completable Sublanguage and Its Computation}



To solve the BTNSCP formulated in Section 3,
we first present a basic result which is a counterpart to Theorem~\ref{thm:sct} in (\cite{BraWon94,Wonham16a}) and Theorem 6 in \cite{ZhangCaiAutomatica:2024}.

\smallskip
\begin{thm} \label{thm:snsct}
Consider a TDES plant modelled by $tick$-automaton ${\bf G} = (Q, \Sigma_c \dot\cup \Sigma_{uc},\delta,q_0,Q_m)$,
{a cover $\mathcal{Q}_{\bf G} = $ $\{Q_{m,i} \subseteq Q_m | i \in \mathcal{I}\}$ on $Q_m$, and a positive integer $N_i$ associated with each $Q_{m,i} \in \mathcal{Q}_{\bf G}$.}
Let $K\subseteq L_m({\bf G})$, $K \neq \emptyset$. There exists a bounded-time nonblocking (marking) TDES supervisory control $V$ (for $(K, {\bf G})$) such that $L_m(V/{\bf G}) = K$ if and only if $K$ is controllable and bounded-time completable wrt. $\{({Q_{m,i}}, N_i)|i \in \mathcal{I}\}$. Moreover, if such a bounded-time nonblocking TDES supervisory control $V$ exists, then it may be implemented by a bounded-time nonblocking $tick$-automaton ${\bf QSUP}$, i.e. $L_m({\bf QSUP}) = L_m(V/{\bf G})$.
\hfill $\diamond$
\end{thm}

Theorem~\ref{thm:snsct} asserts that when the $K$-synthesizing supervisory control $V$ is required to be bounded-time nonblocking, it is necessary and sufficient to require that $K$ be not only controllable but also bounded-time completable.
If $K$ is indeed controllable and bounded-time completable, then the TDES supervisory control $V$ in Theorem~\ref{thm:snsct} is the solution to the BTNSCP. If $K$ is either not controllable or not bounded-time completable, then to achieve the third requirement of maximal permissiveness of BTNSCP, one hopes that the supremal controllable and bounded-time completable sublanguage of $K$ exists. The key is to investigate if for bounded-time completability the supremal element also exists. We provide a positive answer below. Before we proceed, the following is a proof of Theorem~\ref{thm:snsct}.

{\it Proof of Theorem~\ref{thm:snsct}.}
We first prove the first statement. The direction of (only if) is a direct result from Theorem~\ref{thm:sct} and Definition~\ref{defn:snnbs}.
For the direction of (if), according to Theorem~\ref{thm:sct}, since $K$ is controllable, there exists a TDES supervisory control $V$ such that
$V$ is nonblocking and $L_m(V/{\bf G}) = K$. Furthermore, according to Definition~\ref{defn:snnbs}, it is derived from $L_m(V/{\bf G}) = K$ being bounded-time completable wrt. $\{(Q_{m,i}, N_i)|i \in \mathcal{I}\}$ that  $V$ is bounded-time nonblocking wrt. $\{({Q_{m,i}}, N_i)|i \in \mathcal{I}\}$.

For the second statement, let $V$ be a bounded-time nonblocking supervisory control that synthesizes a controllable and bounded-time completable $K$, i.e. $L_m(V/{\bf G}) = K$.
Since $K$ is bounded-time completable,
it follows from Proposition~\ref{prop:relation} that there exists a bounded-time nonblocking $tick$-automaton ${\bf QSUP}$ such that $L_m({\bf QSUP}) = K = L_m(V / {\bf G})$. This completes the proof.
\hfill $\blacksquare$

\subsection{Supremal Bounded-Time Completable Sublanguage}  \label{subsec:supremal}

Let ${\bf G}$ be a nonblocking $tick$-automaton. First, we present the following proposition that bounded-time language completability is closed under arbitrary set unions.

\begin{pro} \label{pro:sncomp}
Consider a $tick$-automaton ${\bf G} = $ $(Q, \Sigma,\delta,q_0,Q_m)$, a cover $\mathcal{Q}_{\bf G} = \{Q_{m,i} \subseteq Q_m | i \in \mathcal{I}\}$ on $Q_m$, and a positive integer $N_i$ associated with each $Q_{m,i} \in \mathcal{Q}_{\bf G}$.
Let $K_1, K_2 \subseteq L_m({\bf G})$. If both $K_1$ and $K_2$ are bounded-time completable wrt. $\{({Q_{m,i}}, N_i)|i \in \mathcal{I}\}$,
then $K := K_1\cup K_2$ is also bounded-time completable wrt. $\{({Q_{m,i}}, N_i)|i \in \mathcal{I}\}$.
\end{pro}

{\it Proof:}
Let $s \in \overline{K}$ and {$i \in \mathcal{I}$.}
According to Definition~\ref{defn:sncomp}, to show that $K$ is bounded-time completable, we need to show that (i) $M_{K,i}(s) \neq \emptyset$, i.e. there exists $t \in \Sigma^*$ such that {$st \in K_i = K \cap L_{m,i}({\bf G})$}, and (ii) for all {$t \in M_{K,i}(s)$}, $\#t(tick) \leq N_i$. Since $\overline{K} = \overline{K_1 \cup K_2} = \overline{K_1} \cup \overline{K_2}$, either $s \in \overline{K}_1$ or $s \in \overline{K}_2$. We consider the case $s \in \overline{K}_1$; the other case is similar.

We first show that (i) holds. Since $K_1$ is bounded-time completable, $M_{K_1,i}(s) \neq \emptyset$, i.e. there exists
string $t$ such that $st \in K_1 \cap L_{m,i}({\bf G}) \subseteq K \cap L_{m,i}({\bf G})$. Thus (i) is established.

For (ii), let $t \in M_{K,i}(s)$; then $st \in K\cap L_{m,i}({\bf G})$ and for all $t'\in \overline{t} \setminus\{t\}$, $st' \notin K\cap L_{m,i}({\bf G})$.
Since $K = K_1 \cup K_2$, there exist the following two cases: (a) $st \in K_1 \cap L_{m,i}({\bf G}) $ and for all $t'\in \overline{t} \setminus\{t\}$, $st' \notin K\cap L_{m,i}({\bf G})$;
(b) $st \in K_2 \cap L_{m,i}({\bf G})$ and for all $t'\in \overline{t} \setminus\{t\}$, $st' \notin K \cap L_{m,i}({\bf G})$.
For case (a), it follows from $K \supseteq K_1$ that $st' \notin K_1\cap L_{m,i}({\bf G})$, so $t \in M_{K_1,i}(s)$. Since $K_1$ is bounded-time completable, it holds that $\#t(tick) \leq N_i$.
The same conclusion holds for case (b) by a similar argument on $K_2$. Hence (ii) is established.

With (i) and (ii) as shown above, we conclude that $K$ is bounded-time completable.

~~~~~~~~~~~~~~~~~~~~ \hfill $\blacksquare$


Now for a given sublanguage $K \subseteq L_m({\bf G})$, whether or not $K$ is bounded-time completable wrt. $\{({{Q}_{m,i}}, N_i)|i \in \mathcal{I}\}$, let
\begin{align*}
{\mathcal{BTC}}(K, \{({ Q_{m,i}}, N_i) |i \in \mathcal{I}\}):=\{K' \subseteq K \mid K' \text{is bounded-time completable} \notag\\
 \text{wrt.}\{({Q_{m,i}}, N_i)|i \in \mathcal{I}\}\} \notag\\
\end{align*}
represent the set of sublanguages of $K$ that are bounded-time completable wrt. $\{({{Q}_{m,i}}, N_i)|i \in \mathcal{I}\}$.
Note from Definition \ref{defn:sncomp} that the empty language $\emptyset$ is trivially bounded-time completable, so $\emptyset \in $ ${\mathcal{BTC}}(K, \{({ Q_{m,i}}, N_i) | i \in \mathcal{I}\})$ always holds.
Moreover, it follows from Proposition~\ref{pro:sncomp} that there exists the supremal bounded-time completable sublanguage of $K$ wrt. $\{({{Q}_{m,i}}, N_i)|i \in \mathcal{I}\}$, given by
\begin{align*}
\sup\mathcal{BTC}(K, &\{({ Q_{m,i}}, N_i) | i \in \mathcal{I}\}) := \bigcup\{ K' \mid  K' \in \mathcal{BTC}(K, \{({ Q_{m,i}}, N_i) | i \in \mathcal{I}\}) \}.
\end{align*}

To compute this $\sup\mathcal{BTC}(K, \{({ Q_{m,i}}, N_i) | i \in \mathcal{I}\})$, we proceed as follows.
Fix $i \in \mathcal{I}$ and let
\begin{align*}
{\mathcal{BTC}}&(K, ({ Q_{m,i}}, N_i)):= \{K' \subseteq K \mid K'~ \text{is bounded-time completable wrt.}({Q_{m,i}}, N_i)\}
\end{align*}
be the set of all bounded-time completable sublanguage of $K$ wrt. $({Q_{m,i}}, N_i)$ (Definition \ref{defn:sncomp}).
By the same reasoning as above, we have that
$\sup\mathcal{BTC}(K, ({Q_{m,i}}, N_i))$ exists. The idea of our algorithm design is to first compute $\sup\mathcal{BTC}(K, ({ Q_{m,i}}, N_i))$ for a fixed $i \in \mathcal{I}$, and then iterate over all $i \in \mathcal{I}$ until fixpoint in order to compute $\sup\mathcal{BTC}(K, \{({Q_{m,i}}, N_i) | i \in \mathcal{I}\})$.
In the next subsection, we present an automaton-based algorithm to compute $\sup\mathcal{BTC}(K, (Q_{m,i}, N_i))$ for any given language $K \subseteq L_m({\bf G})$.

\subsection{Automaton-Based Computation of $\sup\mathcal{BTC}(K, ({Q_{m,i}}, N_i))$} \label{subsec:computation}

Consider a language $K \subseteq L_m({\bf G})$ and $({Q_{m,i}}, N_i)$ for a fixed $i \in \mathcal{I}$.
We present an algorithm (adapted from Algorithm 1 in \cite{ZhangCaiAutomatica:2024} for computing the supremal quantitatively completable sublanguage)
to compute the supremal bounded-time completable sublanguage  $\sup\mathcal{BTC}(K, ({Q_{m,i}}, N_i))$.
The intuition is that we find for each prefix of $\overline{K}$ the bounded-time completable strings, and remove
other non-bounded-time completable strings from the transition graph. The detailed steps are described in Algorithm~\ref{algm:supnc}. In the algorithm,
we employ a last-in-first-out stack $ST$ to store the states to be processed (a first-in-first-out queue can also be used instead to perform a different order of search), and for a set $Z$ a flag $F:Z \rightarrow \{true, false\}$ to
indicate whether or not an element of $Z$ has been visited: i.e. $F(z) = true$ iff $z \in Z$ has been visited.

\begin{algorithm}[!h]
\caption{: Algorithm of Computing $\sup\mathcal{BTC}(K,(Q_{m,i}, $  $N_i))$}
\label{algm:supnc}

\noindent {\bf Input}: $tick$-automaton ${\bf G} = (Q, \Sigma, \delta, q_0, Q_m)$, language $K \subseteq L_m({\bf G})$, subset $Q_{m,i} \subseteq Q_m$ of marker states, and positive integer $N_i$.\\
\noindent {\bf Output}: Language $K_i'$.
\\
\\
\noindent {\bf Step 1.} Construct a $tick$-automaton ${\bf K}_i = (X_i, \Sigma, \xi_i, x_{i,0},$ $ X_{i,m})$ to represent $K_i = K \cap L_{m,i}({\bf G})$. If $K_i = \emptyset$, output language $K_i' = \emptyset$; otherwise go to Step 2.\\
\\
\noindent {\bf Step 2.} Let
\[X_i' := \{(x_i,d)|x_i\in X_i, d \in \{0,...,N_i\}\},\]
$\xi_i' = \emptyset$, $x_{i,0}' = (x_{i,0},0)$, and $X_{i,m}' := \{(x_i,0)|x_i\in X_{i,m}\}$. Initially set $F((x_i,d)) = false$ for each state $x_i \in X_i$ and each $d \in \{0,...,N_i-1\}$.
Then push the initial state $x_{i,0}'=(x_{i,0},0)$ into stack $ST$, and set $F((x_{i,0},0)) = true$.
\\
\\
\noindent {\bf Step 3.} If stack $ST$ is empty, trim\footnotemark ~the $tick$-automaton ${\bf K}_i' = (X_i', \Sigma, \xi_i', x'_{i,0}, X_{i,m}')$,
and go to Step 6.
Otherwise, pop out the top element $(x_{i,j}, d)$ of stack $ST$. If $x_{i,j} \in X_{i,m}$, go to Step 4; otherwise, go to Step 5.
\\
\\
\noindent {\bf Step 4.} For each event $\sigma \in \Sigma$ defined at state $x_{i,j}$ (i.e. {$\xi_i(x_{i,j},\sigma)!$}), let $x_{i,k} := {\xi_i(x_{i,j},\sigma)}$ and do the following two steps 4.1 and 4.2; then go to Step 3 with updated stack $ST$.
\\
\indent~~ {\bf Step 4.1} Add transition $((x_{i,j},0),\sigma,(x_{i,k},0))$ to {$\xi_i'$}, i.e.
\[\xi_i' := \xi_i' \cup \{((x_{i,j},0),\sigma, (x_{i,k},0))\}.\]
\indent~~ {\bf Step 4.2} If $F((x_{i,k},0)) = false$, push $(x_{i,k},0)$ into stack $ST$ and set $F((x_{i,k},0)) = true$.
\\
\\
\noindent {\bf Step 5.} For each event $\sigma \in \Sigma$ defined at state $x_{i,j}$ (i.e. {$\xi_i(x_{i,j},\sigma)!$}), do the following three steps 5.1--5.3;
then go to Step 3 with updated stack $ST$.
\\
\indent~~ {\bf Step 5.1} Let $x_{i,k} := \xi_i(x_{i,j},\sigma)$. If $\sigma = tick$, set $d' = d + 1$;
If $\sigma \neq tick$ and $x_{i,k} \in X_{i,m}$, set $d' = 0$; if $\sigma \neq tick$ and $x_{i,k} \notin X_{i,m}$, set $d' = d$.
\\
\indent~~ {\bf Step 5.2} If $d' > N_i$, go to Step 5.1 with the next event $\sigma$ defined at $x_{i,j}$. Otherwise, add a new transition $((x_{i,j},d),\sigma, (x_{i,k},d'))$ to {$\xi_i'$}, i.e.
\[\xi_i' := \xi_i' \cup \{((x_{i,j},d),\sigma, (x_{i,k},d'))\}\]
\indent~~ {\bf Step 5.3} If $F((x_{i,k},d')) = false$, push $(x_{i,k},d')$ into stack $ST$ and set $F((x_{i,k},d')) = true$.
\\
\\
\noindent {\bf Step 6.} Output the language $K_i' = K \cap L({\bf K}_i')$.
\\
\end{algorithm}
\footnotetext{`Trimmed' means that all non-reachable and non-coreachable states (if they exist) are removed (\cite{Elienberg:1974, Wonham16a}).
The $tick$-automaton ${\bf K}'$ need not be trim, and to get a nonblocking $tick$-automaton, this step of trimming is required.  }

In Step 5.2 of Algorithm~\ref{algm:supnc} above, note that the condition $d' > N_i$ means that the downstream transitions including more than $N_i$ $ticks$ that have never reached a marker state in $Q_{m,i}$ will be removed, therefore guaranteeing that from an arbitrary state, at most $N_i$ $ticks$
are needed to reach a marker state in $Q_{m,i}$.

Note that there are two differences between the above new Algorithm~\ref{algm:supnc} and the algorithm in \cite{ZhangCaiAutomatica:2024} for computing supremal quantitatively completable language. The first difference is the definition of $X_i'$. In the new algorithm, we have
$X_i' := \{(x_i,d)|x_i\in X_i, d \in \{0,...,N_i\}\}$, namely, $d$ is in the range of $[0, N_i]$, rather than $[0,N_i-1]$
in the Algorithm 1 in \cite{ZhangCaiAutomatica:2024}. The reason is illustrated in Fig.~\ref{fig:DN}.
\begin{figure}[!t]
\centering
    \includegraphics[scale = 0.4]{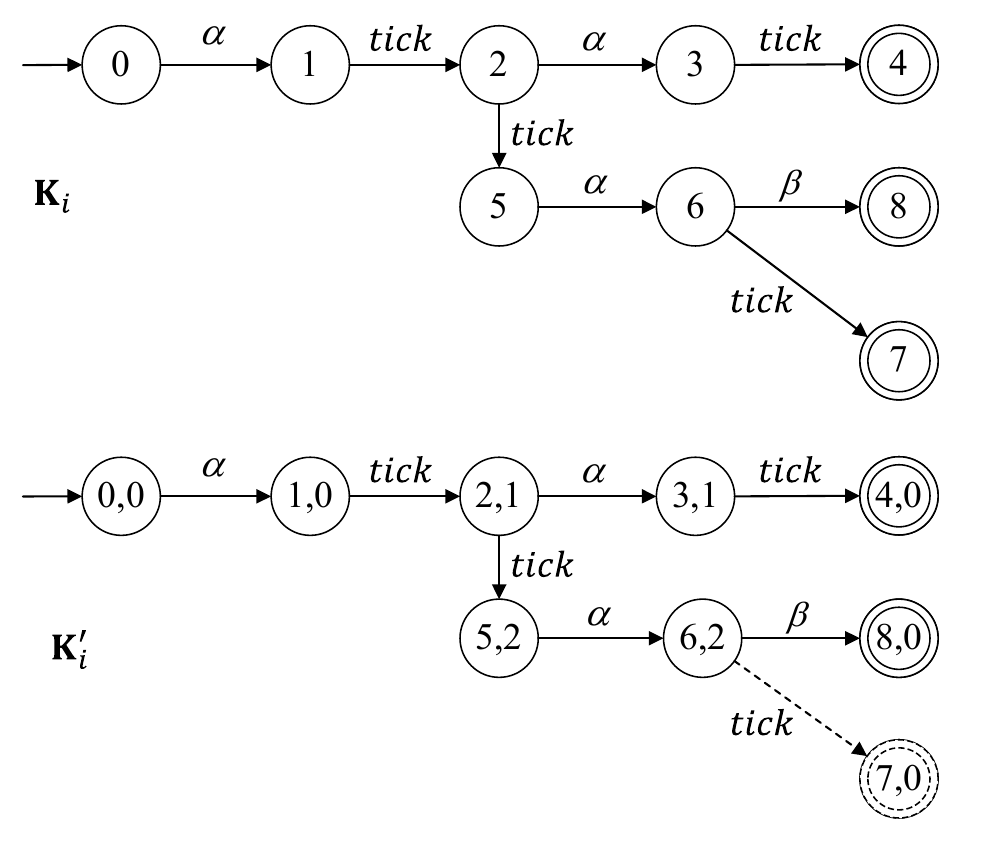}
\caption{Illustration on the reason for defining $d \in [0, N_i]$ in $X_i'$. Assume that the top subfigure represents partial transition functions of a $tick$-automaton ${\bf K}_i$ and let $N_i = 2$. In the process of constructing ${\bf K}_i'$ (as represented by the bottom subfigure) by Algorithm~\ref{algm:supnc}, from state (2,1) to state (5,2), $d = 1$ is advanced to $d' = 2$ since $\sigma = tick$. In the algorithm in \cite{ZhangCaiAutomatica:2024} of computing
the supremal quantitatively completable sublanguage, since state 5 is not a marker state, this transition will be removed. However, since the downstream string $\alpha.\beta$ will lead the automaton to marker state 8 (where $d'$ will be set to 0) and the total number of occurred $ticks$ is no more than 2 $ticks$, such states and transitions should be reserved. Thus in Algorithm~\ref{algm:supnc}, by Step 5.2 the states with $d' = N_i$ are included in ${\bf K}_i'$. Also by this step, the states with $d'$ larger than 2 are removed from ${\bf K}_i$; e.g., transition $(6, tick, 7)$ is removed by Step 5.2 because now $d' = 3 > N_i$ (as represented by the dashed lines in the transition graph of ${\bf K}_i'$). } \label{fig:DN}
\end{figure}

The second difference is the rules for updating $d$ at Step 5.1: in Algorithm 1 above, only when $\sigma = tick$ will $d'$ be set to $d+1$. The reason is that only the $tick$ transitions are counted for satisfying bounded-time completability; while in Algorithm 1 in \cite{ZhangWangCai:2021}, the transitions caused by all events are counted. Namely, in the above new algorithm,
the $tick$ and non-$tick$ events on the transitions need be distinguished for selecting different rules for updating $d$. However, in the algorithm in \cite{ZhangWangCai:2021},
no such event distinguishing is needed.


Now we present an example to illustrate Algorithm~\ref{algm:supnc}.

\begin{example} [Continuing Example 2.2]
Inputting $tick$-automaton ${\bf G}$, language $K = L_m({\bf SUP})$, marker state subset ${Q_{m,1}} = {\{4,11\}} \subseteq Q_m$ and
positive integer $N_1 = 5$, Algorithm~\ref{algm:supnc} generates a
trimmed automaton ${\bf TSUP}_1$, as displayed in Fig.~\ref{fig:TSUP1}.

From Fig.~\ref{fig:TSUP1}, we observe that the the algorithm will terminate the search
before the sixth $tick$ occurs. Every reachable states are guaranteed to
reach the marker states (4,0) and (11,0) in at most 5 $ticks$.
By contrast in ${\bf SUP}$ (displayed in Fig.~\ref{fig:SUP}), the strings $(tick)^{5}.1.tick.2$ will lead state 0 to the marker state (4,0); however, they
include more than 5 $ticks$. Thus, in fact those strings are removed from ${\bf SUP}$ by Algorithm~\ref{algm:supnc}. Also, the states and the transitions that cannot reach marker state are removed by the trim operation at Step 3. \hfill $\diamond$

\begin{figure*}[!t]
\centering
    \includegraphics[scale = 0.34]{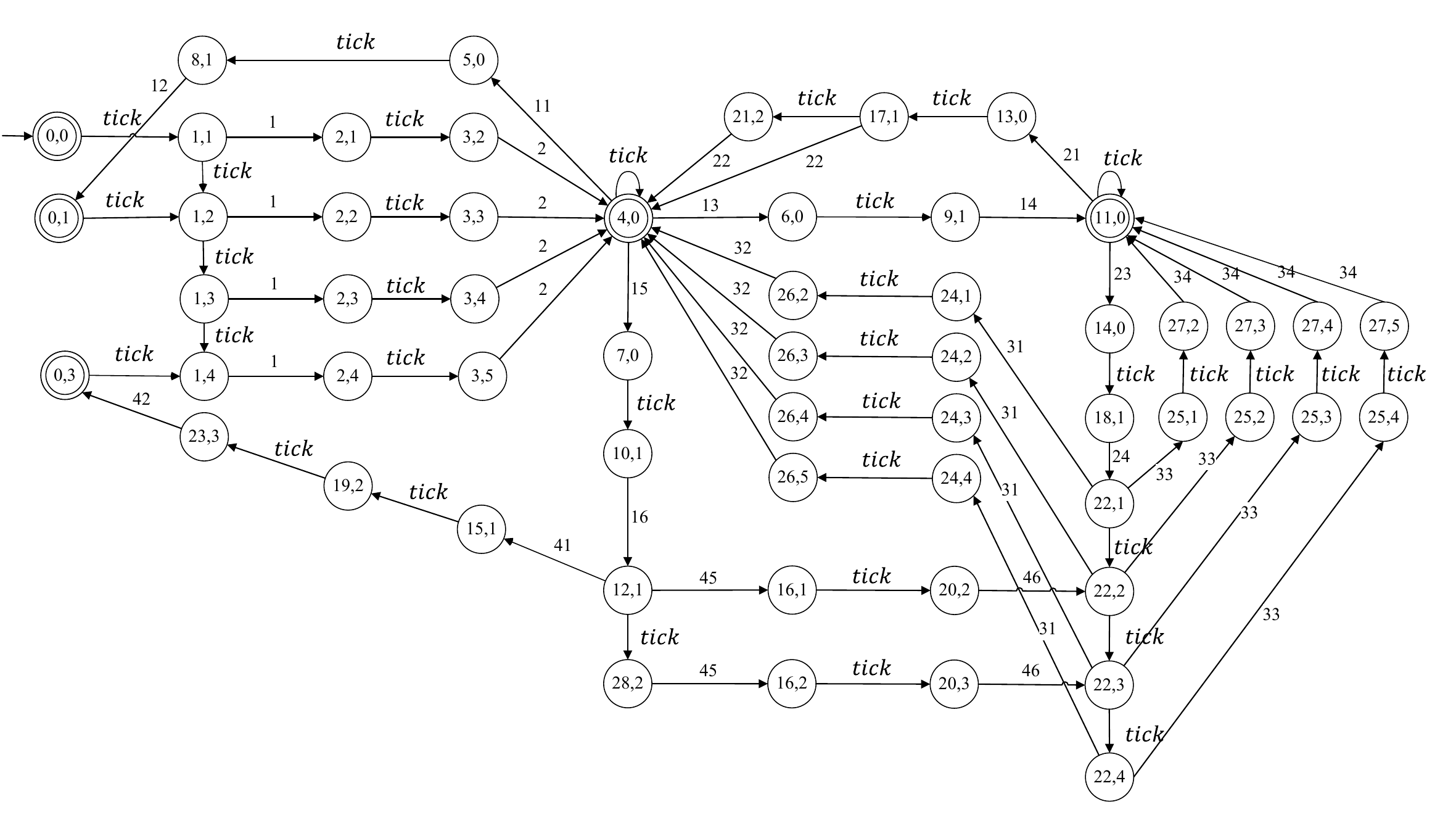}
\caption{Transition graph of ${\bf TSUP}_1$} \label{fig:TSUP1}
\end{figure*}
\end{example}

The correctness of Algorithm 1 is confirmed by the following theorem.

\begin{thm}
Given a $tick$-automaton $\bf G$, a sublanguage $K \subseteq L_m({\bf G})$, subset $Q_{m,i} \subseteq Q_m$ of marker states, and positive integer $N_i$,
let $K_i'$ be the language returned by Algorithm~\ref{algm:supnc}. Then
$K_i' = \sup\mathcal{{BTC}}(K, ({Q_{m,i}}, N_i))$.
\end{thm}

{\it Proof:} First, we prove that $K_i' \in \mathcal{{BTC}}(K, ({Q_{m,i}}, N_i))$.
We start by showing that $K_i' \subseteq K$, which can be directly obtained by Step 6.

Next we show that $K_i'$ is bounded-time completable wrt. $(Q_{m,i}, N_i)$. For this, let ${\bf K}_i' = (X_i', \Sigma, \xi_i', x_{i,0}',$ $ X_{i,m}')$
be the $tick$-automaton generated by Step 3 of Algorithm ~\ref{algm:supnc}. According to Step 6, it is easily verified
that $\overline{K_i'} = L({\bf K}_i')$ and $K_i' \cap L_{m,i}({\bf G}) = L_m({\bf K}_i') \cap L_{m,i}({\bf G})$ (because $K_i' \cap L_{m,i}({\bf G}) =
K \cap L({\bf K}_i') \cap L_{m,i}({\bf G}) = L_m({\bf K}_i) \cap L({\bf K}_i') \cap L_{m,i}({\bf G}) = L_m({\bf K}_i') \cap L_{m,i}({\bf G})$).
Let $s \in \overline{K_i}'$, then $s \in \overline{K}$ and $s \in L({\bf K}_i')$; then there exists $t \in \Sigma^*$ such that $st\in L_m({\bf K}_i')$,
i.e. $st \in K_i' \cap L_{m,i}({\bf G})$. According to the definition of $M_{K_i',i}(s)$, $M_{K_i',i}(s) \neq \emptyset$; namely, condition
(i) of bounded-time completability is satisfied.
{Furthermore, according to the definition of $M$ in (\ref{eq:PKL}), it is derived from $K_i' \cap L_{m,i}({\bf G}) = L_m({\bf K}_i') \cap L_{m,i}({\bf G})$ that for each $t \in M_{K_i',i}(s)$, $t \in M_{L_m({\bf K}_i'),i}(s)$}. It is guaranteed by  $d' > N_i$ in Step 5.2 that if a string includes more than $N_i$ $ticks$ and has not reached a marker state in $Q_{m,i}$, it will not be added to $L({\bf K}_i')$. In other words, those strings $st$ added to $L({\bf K}_i')$ must satisfy
that for every $t \in M_{L_m({\bf K}_i'),i}(s)$, there holds $\#t(tick) \leq N_i$. Namely, condition (ii) of bounded-time completability
is also satisfied. Hence, $K_i'$ is bounded-time completable wrt. $(Q_{m,i}, N_i)$.  This establishes that $K_i' \in \mathcal{BTC}(K, (Q_{m,i}, N_i))$.

It remains to show that $K_i'$ is the largest element in $\mathcal{BTC}(K, (Q_{m,i}, N_i))$.
Let $M$ be another element in $\mathcal{BTC}(K, (Q_{m,i}, N_i))$, i.e. $M \in \mathcal{BTC}(K, (Q_{m,i}, N_i))$. It will be shown that  $M \subseteq K_i'$. For this, we first prove that $\overline {M} \subseteq \overline{K_i'} = L({\bf K}_i')$ by induction on the length of a string $s \in \overline{M}$.

{\bf Base case:} Let $s = \epsilon \in \overline{M}$. Then $\epsilon \in \overline{K}$ and the initial state $x_0$ exists in ${\bf K}_i$. It follows from Step~1 that $x_{i,0}' = (x_{i,0},0)$ is designated to be the initial state of ${\bf K}_i'$, and hence $\epsilon \in L({\bf K}_i')$.

{\bf Inductive case:} Let $s \in \overline{M}$, $s \in L({\bf K}_i')$, $\sigma \in \Sigma$, and suppose that $s\sigma \in \overline{M}$; we will show that $s\sigma \in L({\bf K}_i')$ as well.
Since $M \in \mathcal{BTC}(K, (Q_{m,i}, N_i))$, we have (i) $M \subseteq K$ and (ii) $M$ is bounded-time completable wrt $(Q_{m,i}, N_i)$.
It follows from (i) that {$s \in \overline{K} \cap L({\bf G}) = L({\bf K}_i)$}, i.e.
$\xi_i(x_{i,0}, s)!$. By the same reason, $\xi_i(x_{i,0},s\sigma)!$. Letting $x_i = \xi_i(x_{i,0}, s)$, we derive $\xi_i(x_i,\sigma)!$. Since $s \in L({\bf K}_i')$, $\xi_i'((x_{i,0},0),s)!$.
According to the definition of $\xi_i'$, there must exist $d \in \{0,...,N_i\}$ such that $(x_i,d) = \xi_i'((x_{i,0},0),s)$.
We already know that $\xi_i(x_i,\sigma)!$. If $x_i \in X_{i,m}$, according to Step 4, $\xi_i'((x_i,d),\sigma)$ with $d = 0$ is defined.
If $x_i \notin X_{i,m}$, according to Step 5.1, there may exist the following three cases: (a) $\sigma = tick$; (b) $\sigma \neq tick$ and $\xi_i(x_i,\sigma) \in X_{i,m}$; (c) $\sigma \neq tick$ and $\xi_i(x_i,\sigma) \notin X_{i,m}$.
First, in case (a), $d' = d+1$. According to Step 5.2, $d' > N_i$ or $d' \leq N_i$. The former is impossible because in that case $s \notin M$ (since $x_i \notin X_{i,m}$, $s \notin K$ or $s \notin L_{m,i}({\bf G})$; in both cases, $s \notin M$) but $s.tick \in \overline{M}$ will imply that it needs at least $N_i+1$ $ticks$ to lead $\overline{M}$ to $M$, which
contradicts to the assumption that $M$ is bounded-time completable $(Q_{m,i}, N_i)$. When $d' \leq N_i$,
$\xi'((x,d),\sigma)$ is defined with $d' \leq N_i$.
For case (b), $d' = 0$, according to Step 5.2, $\xi'((x,d),\sigma)$ is defined with $d' = 0$.
For case (c), $\xi'((x,d),\sigma)$ is defined with $d \leq N$ and thus $d' = d \leq N_i$. 
Hence, we conclude that $\xi_i'((x_i,d),\sigma)$ is defined, i.e. $s\sigma \in L({\bf K}_i')$.

Therefore, by the above induction, $\overline {M} \subseteq L({\bf K}_i')$ is established.
So, we have
\begin{align*}
M &\subseteq \overline{M} \cap K  ~~~(\text{by}~ M \in \mathcal{BTC}(K, (Q_{m,i}, N_i)))\\
    & \subseteq L({\bf K}_i') \cap K ~~~(\text{according to Step 6})\\
    & \subseteq K_i'.
\end{align*}
The proof is now complete. \hfill $\blacksquare$

The above theorem confirms that Algorithm 1 computes the supremal bounded-time completable sublanguage {$\sup\mathcal{BTC}(K, ({Q_{m,i}}, N_i))$}.
The time complexity of Algorithm~\ref{algm:supnc} is $O(|Q|\cdot|X|\cdot|\Sigma|\cdot N_i)$ where $|Q|$ and $|X|$ are the states number of ${\bf G}$ and $tick$-automaton representing $K$ respectively. This complexity is derived according to Steps 3 to 5, because
${\bf K}_i$ has at most $|Q| \cdot |X| \cdot |\Sigma|$ transitions and each transitions are visited at most $N_i$ times.

\section{Maximally Permissive Bounded-Time Nonblocking Supervisory Control}


In this section, we present our solution to the BTNSCP.
Consider a TDES plant modelled by $tick$-automaton ${\bf G} = (Q,\Sigma_c \dot\cup \Sigma_{uc},\delta,q_0,Q_m)$, and a specification language $E \subseteq \Sigma^*$. Let $K := E \cap L_m({\bf G})$, $\mathcal{Q}_{\bf G} = \{Q_{m,i} | i \in \mathcal{I}\}$ be a cover on $Q_m$, and a positive integer $N_i$ associated with each $Q_{m,i} \in \mathcal{Q}_{\bf G}$.

Whether or not $K$ is controllable and bounded-time completable, let $\mathcal{CBTC}(K, {\{(Q_{m,i}, N_i)| i \in \mathcal{I}\}})$ be the set of sublanguages of $K$ that are both controllable and bounded-time completable wrt. $\{(Q_{m,i}, N_i)| i \in \mathcal{I}\}$, i.e.
\begin{align*}
\mathcal{CBTC}(K, {\{(Q_{m,i}, N_i)| i \in \mathcal{I}\}})& :=\{K' \subseteq K \mid K' \text{is controllable and} \notag\\
& ~\text{bounded-time completable wrt. ${\{(Q_{m,i}, N_i)| i \in \mathcal{I}\}}$}\}\notag
\end{align*}
Since the empty language $\emptyset$ is trivially controllable and bounded-time completable, the set $\mathcal{CBTC}(K, \{(Q_{m,i}, N_i)| i \in \mathcal{I}\})$ is nonempty. Moreover, since both controllability and bounded-time completability are closed under arbitrary set unions, $\mathcal{CBTC}(K, \{(Q_{m,i}, N_i)| i \in \mathcal{I}\}$ contains a unique supremal element given by
\begin{align*}
\sup\mathcal{CBTC}(K, &{\{(Q_{m,i}, N_i)| i \in \mathcal{I}\}}):=\bigcup\{K' \subseteq K \mid K' \in \mathcal{CBTC}(K, {\{(Q_{m,i}, N_i)| i \in \mathcal{I}\}}) \}.
\end{align*}

Our main result in this section is the following.

\begin{thm}
Suppose that $\sup\mathcal{CBTC}(K, {\{(Q_{m,i}, N_i)}|i \in \mathcal{I}\})\neq \emptyset$.
Then the supervisory control $V_{\sup}$ such that $L_m(V_{\sup}/{\bf G})= \sup\mathcal{CBTC}(K, {\{(Q_{m,i}, N_i)| i \in \mathcal{I}\}}) $ $\subseteq K$ is the solution to the BTNSCP.
\end{thm}

The proof of the above result is similar to that of Theorem 10 in \cite{ZhangCaiAutomatica:2024} with $\sup\mathcal{CQC}(K, {\{(Q_{m,i}, N_i)}|i \in \mathcal{I}\})$ replaced by $\sup\mathcal{CBTC}(K, {\{(Q_{m,i}, N_i)}|i \in \mathcal{I}\})$.



We proceed to design an algorithm (adapted from Algorithm 4 in \cite{ZhangCaiAutomatica:2024} for computing the supremal
controllable and quantitatively completable sublanguage) to compute this solution $\sup\mathcal{CBTC}(K, \{(Q_{m,i}, N_i)| i \in \mathcal{I}\})$.
The change is that the algorithm for computing the supremal quantitatively completable sublanguage is replaced by Algorithm 1 in Section 4.2 for
computing the supremal bounded-time completable sublanguage. For self-containedness, we present the algorithm below.

\begin{algorithm}[!h] {
\caption{: Algorithm of Computing $\sup\mathcal{CBTC}(K, \{(Q_{m,i}, N_i)| i \in \mathcal{I}\})$\quad ($\mathcal{I} = \{1,\ldots,M\}$)}
\label{algm:supcsnc}

\noindent {\bf Input}: $tick$-automaton ${\bf G} = (Q, \Sigma, \delta, q_0, Q_m)$, language $K \subseteq L_m({\bf G})$, cover $\mathcal{Q}_{\bf G} = \{Q_{m,i} | i \in \mathcal{I}\}$
on marker state set $Q_m$, and set of positive integers $\{N_i| i \in \mathcal{I}\}$. \\
\noindent {\bf Output}: Language $CTK$.
\\

\noindent {\bf Step 1.} Let $j = 1$ and $K^j = K$ (i.e. $K^1 = K$).

\noindent {\bf Step 2.} Let $i = 1$. Let $K_i^j = K^j$.

\noindent {\bf Step 2.1} Apply Algorithm~\ref{algm:supnc} with inputs ${\bf G}$, $K_i^j$, $Q_{m,i}$ and $N_i$, and obtain ${NK_i^j} = \sup\mathcal{BTC}(K_i^j, (Q_{m,i}, N_i))$.

\noindent {\bf Step 2.2} If $i < M$, let $K_{i+1}^j = {NK_i^j}$, advance $i$ to $i+1$ and go to Step 2.1; otherwise ($i = M$), go to Step 3.

\noindent {\bf Step 3.} Apply Algorithm SC with inputs ${\bf G}$ and ${NK_{M}^j}$ to compute $K^{j+1}$ such that $K^{j+1} = \sup\mathcal{C}({NK_{M}^j})$.

\noindent {\bf Step 4.} If $K^{j+1} = K^j$, output $CTK = K^{j+1}$. Otherwise, advance $j$ to $j+1$ and go to {\bf Step~2}.
}
\end{algorithm}

The correctness of Algorithm~\ref{algm:supcsnc} is confirmed similar to that of Algorithm 4 in \cite{ZhangCaiAutomatica:2024}
(the detailed proof is referred to \cite{ZhangCaiAutomatica:2024}).

\begin{thm}
Given a $tick$-automaton ${\bf G}$, a specification language $E$, let $K := E \cap L_m({\bf G})$, a cover $\mathcal{Q}_{\bf G} = \{Q_{m,i} \subseteq Q_m | i \in \mathcal{I}\}$ on $Q_m$, and a set of positive integer $N_i$ each associated with a $Q_{m,i} \in \mathcal{Q}_{\bf G}$. Then Algorithm~\ref{algm:supcsnc} terminates in a finite number of steps and outputs a language $CTK$ such that $CTK = \sup\mathcal{CBTC}(K, \{(Q_{m,i}, N_i)| i \in \mathcal{I}\})$.
\end{thm}




The complexity of one complete iteration over all $i \in \mathcal{I}$ in Step 2 of Algorithm~\ref{algm:supcsnc} is $O(|Q|\cdot |X| {\color{red}\cdot |\Sigma|} \cdot \prod_{i=1}^{|\mathcal{I}|}{N_i})$.
Since Algorithm SC does not increase the state/transition number of ${NK_{M}^j}$ ($M= |\mathcal{I}|$), the complexity of each iteration including Steps 2 and 3 is again $O(|Q|\cdot |X| {\color{red}\cdot |\Sigma|}\cdot \prod_{i=1}^{M}{N_i})$.
Finally since there can be at most $|Q|\cdot |X| \cdot |\Sigma| \cdot \prod_{i=1}^M{N_i}$
iterations, the overall time complexity of Algorithm~\ref{algm:supcsnc} is $O\big((|Q|\cdot |X| \cdot |\Sigma| \cdot \prod_{i=1}^M{N_i})^2\big)$.

The following example demonstrates how to synthesize supervisors satisfying
both controllability and bounded-time completability.

\begin{example}[Continuing Example 2.2]
Consider TDES plant modelled by $tick$-automaton ${\bf G}$ and nonblocking supervisor ${\bf SUP}$ displayed in Fig.~\ref{fig:SUP}.
First, applying Algorithm1 with inputs ${\bf G}$, $K = L_m({\bf SUP})$,  $\mathcal{Q}_{\bf G} = \{Q_{m,1} = {\color{red}\{4,11\}}, Q_{m,2} = \{0\}\}$ and $\{N_1 = 5,$ $ N_2 = 9\}$, we get the supremal bounded-time completable language $NK$ (wrt. $\{(Q_{m,1} = {\color{red}\{4,11\}}, N_1 = 5)$, $(Q_{m,2} = \{0\}, N_2 = 9)\}$) represented by $tick$-automaton $\bf NK$ displayed in Fig.~\ref{fig:TQCSUP}.
However, it is not controllable, because event $tick$ is disabled at state 102, but event 22 is not forcible, which cannot preempt the occurrence of $tick$. Next applying Algorithm~SC
with inputs $\bf G$ and $NK$ (although state 83 is a marker state, $tick$ is preempted on it but after event 21 is disabled, none of the forcible
events is eligible at state 83; thus states 83, 94, 95, 98, 99 and corresponding transition arriving them are removed from $\bf NK$), we get an automaton ${\bf TQCSUP}$ as displayed Fig.~\ref{fig:TQCSUP1}.

It is verified that $L_m({\bf TQCSUP})$ is both controllable and bounded-time completable wrt. $\{(Q_{m,1} = {\color{red}\{4,11\}}, N_1 = 5)$, $(Q_{m,2} = \{0\}, N_2 = 9)\}$,
and thus according to Theorem~\ref{thm:snsct}, ${\bf TQCSUP}$ may be used as a bounded-time nonblocking supervisor.

\begin{figure*}[!t]
\centering
    \includegraphics[scale = 0.23]{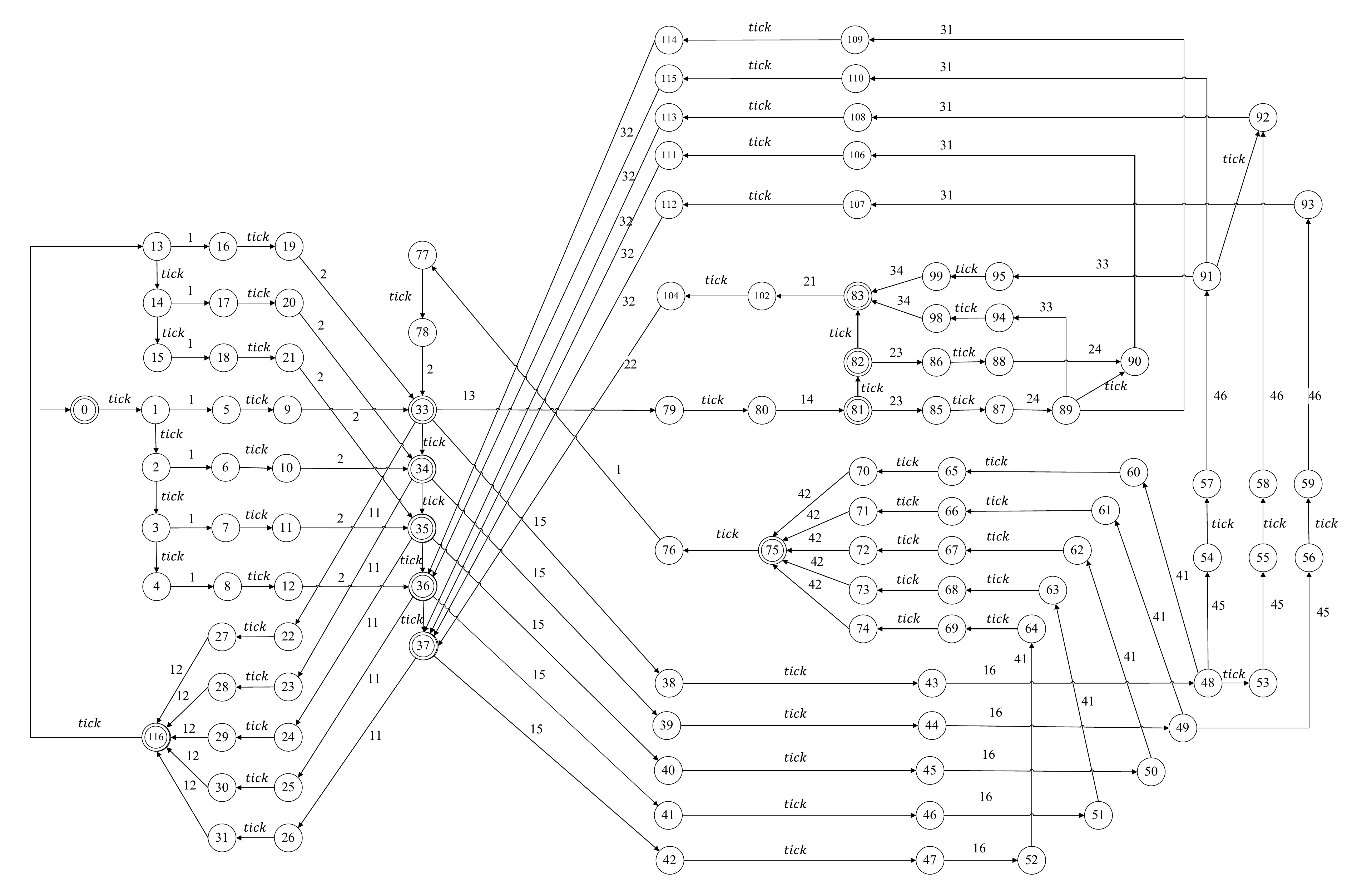}
\caption{Transition graph of ${\bf NK}$ } \label{fig:TQCSUP}
\end{figure*}

This supervisor ${\bf TQCSUP}$ is used to make the autonomous vehicle provide timely services in Example 1.
The control logics of ${\bf TQCSUP}$ are as follows: (1) never move to zone~2 (corresponding to states 81 and 88) and zone~4 (corresponding to
state 48, 49, 50, 51 and 52)  when in zone~3 (corresponding to states 89, 90, 91 and 92); (ii) never move to zone~1 (corresponding to states 33, 34, 35, 35 and 37) when in zone~4; (iii)
if the vehicle is in zone 1, it is safe to move to zone 2 and zone 4 if it has just returned from zone 0 (i.e. finished self-charging; corresponding to states 0, 75, 116);
and (iv) if the vehicle has moved to zone 2 and 3, it must return (either by moving through zone 1, or moving though zone 4) for self-charging before the next round of service.

These logics guarantee that the two requirements ((i) and (ii) in Example~2.2 of Section 2.2) on the vehicle are satisfied.
First, every package sent to customers can be delivered by the vehicle to one of the two service areas (zone~1 or 2) within 10 minutes; and whenever a customer calls for package collection, the vehicle can reach either zone~1 or 2 within 10 minutes no matter where the vehicle is and no matter which paths (permitted by the supervisor $\bf TQCSUP$) the vehicle follows.
Second, no matter where the vehicle is, it can return to zone 0 for self-charging within 18 minutes no matter which paths the vehicle follows.

\begin{figure*}[!t]
\centering
    \includegraphics[scale = 0.23]{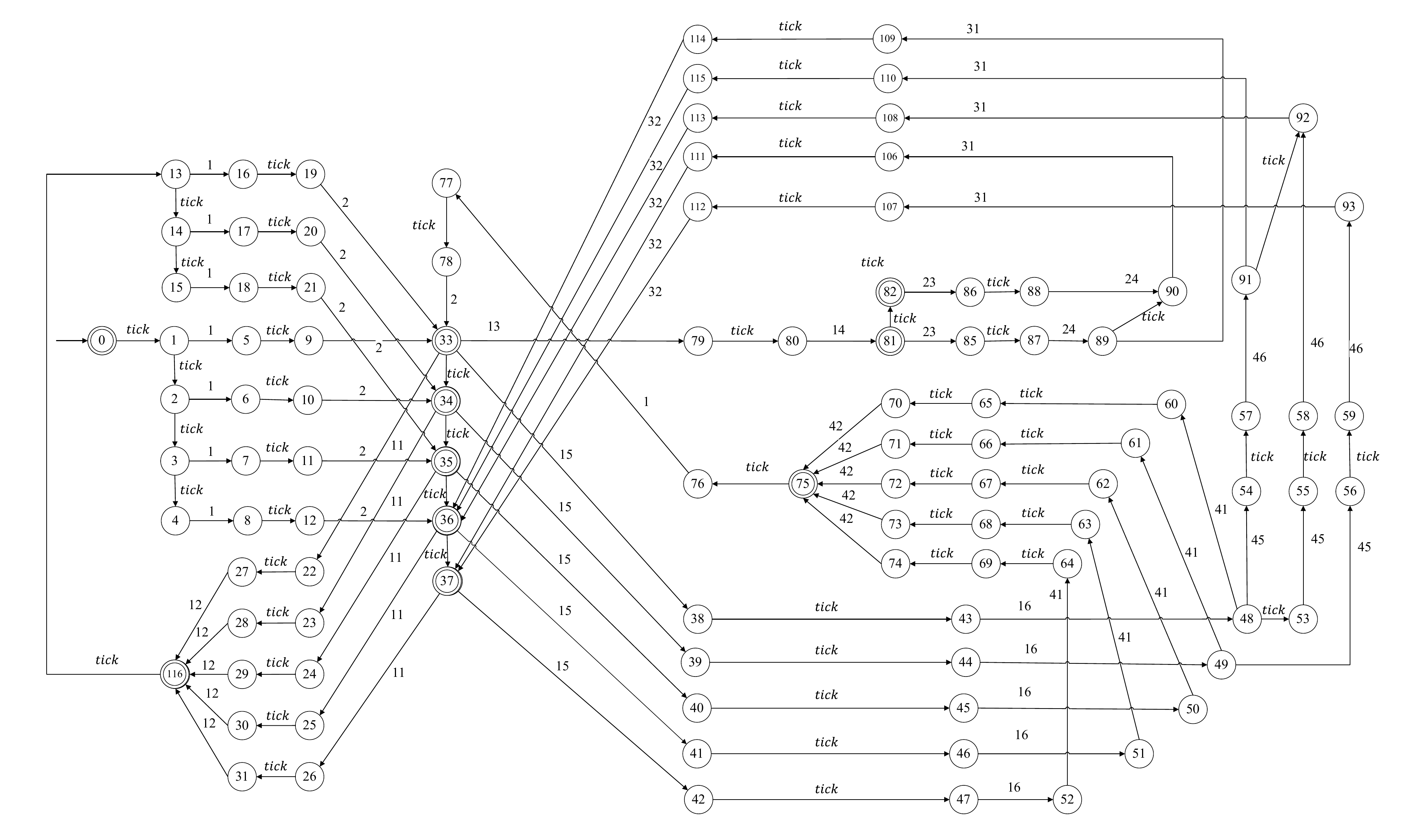}
\caption{Transition graph of ${\bf TQCSUP}$ } \label{fig:TQCSUP1}
\end{figure*}

\end{example}

\section{Conclusion and Future Work}

In this paper, we have introduced a new concept of
bounded-time nonblockingness of $tick$-automaton, which
requires that every task (each represented by a subset
of marker states) must be completed in prescribed
time. Moreover, we
have formulated a new bounded-time nonblocking
supervisory control problem of TDES, characterized its solution
in terms of bounded-time language completability, and
developed algorithms to compute the optimal solution.


\small

\bibliographystyle{IEEEtran}
\bibliography{SCDES_Ref}

\begin{thebibliography}{10}
\providecommand{\url}[1]{#1}
\csname url@samestyle\endcsname
\providecommand{\newblock}{\relax}
\providecommand{\bibinfo}[2]{#2}
\providecommand{\BIBentrySTDinterwordspacing}{\spaceskip=0pt\relax}
\providecommand{\BIBentryALTinterwordstretchfactor}{4}
\providecommand{\BIBentryALTinterwordspacing}{\spaceskip=\fontdimen2\font plus
\BIBentryALTinterwordstretchfactor\fontdimen3\font minus
  \fontdimen4\font\relax}
\providecommand{\BIBforeignlanguage}[2]{{%
\expandafter\ifx\csname l@#1\endcsname\relax
\typeout{** WARNING: IEEEtran.bst: No hyphenation pattern has been}%
\typeout{** loaded for the language `#1'. Using the pattern for}%
\typeout{** the default language instead.}%
\else
\language=\csname l@#1\endcsname
\fi
#2}}
\providecommand{\BIBdecl}{\relax}
\BIBdecl

\bibitem{RamWon87}
P.~Ramadge and W.~Wonham, ``Supervisory control of a class of discrete event
  processes,'' \emph{SIAM Journal on Control and Optimization}, vol.~25, no.~1,
  pp. 206--230, 1987.

\bibitem{WonRam87}
W.~Wonham and P.~Ramadge, ``On the supremal controllable sublanguage of a given
  language,'' \emph{SIAM Journal on Control and Optimization}, vol.~25, no.~3,
  pp. 637--659, 1987.

\bibitem{RamWon89}
P.~Ramadge and W.~Wonham, ``The control of discrete event systems,'' \emph{The
  Proceedings of IEEE}, vol.~77, no.~1, pp. 81--98, 1989.

\bibitem{Wonham16a}
W.~Wonham and K.~Cai, \emph{Supervisory Control of Discrete-Event
  Systems}.\hskip 1em plus 0.5em minus 0.4em\relax Springer, 2019.

\bibitem{CaiWon20}
K.~Cai and W.~Wonham, \emph{Supervisory control of discrete-event
  systems}.\hskip 1em plus 0.5em minus 0.4em\relax Encyclopedia of Systems and
  Control, 2nd ed., Springer, 2020.

\bibitem{WonCaiRud18}
W.~Wonham, K.~Cai, and K.~Rudie, ``Supervisory control of discrete-event
  systems: a brief history,'' \emph{Annual Reviews in Control}, vol.~45, pp.
  250--256, 2018.

\bibitem{CasLaf08}
C.~Cassandras and S.~Lafortune, \emph{Introduction to Discrete Event Systems},
  2nd~ed.\hskip 1em plus 0.5em minus 0.4em\relax Springer, 2008.

\bibitem{BraWon94}
B.~Brandin and W.~Wonham, ``Supervisory control of timed discrete-event
  systems,'' \emph{IEEE Transactions on Automatic Control}, vol.~39, no.~2, pp.
  329--342, 1994.

\bibitem{KumarShayman1994}
R.~Kumar and M.~Shayman, ``Non-blocking supervisory control of deterministic
  discrete event systems,'' in \emph{Proc. 1994 American Control Conference},
  1994, pp. 1089--1093.

\bibitem{FabianKumar:1997}
M.~Fabian and R.~Kumar, ``Mutually nonblocking supervisory control of discrete
  event systems,'' in \emph{Proc. 36th IEEE Conference on Decision and
  Control}, 1997, pp. 2970--2975.

\bibitem{MaWonham:2005}
C.~Ma and W.~Wonham, ``Nonblocking supervisory control of state tree
  structures,'' \emph{IEEE Transactions on Automatic Control}, vol.~51, no.~5,
  pp. 782--793, 2006.

\bibitem{MalikLudec:2008}
R.~Malik and R.~Leduc, ``Generalised nonblocking,'' in \emph{Proc. 9th
  International Workshop on Discrete Event Systems}, 2008, pp. 340--345.

\bibitem{BalemiEt:1993}
S.~Balemi, G.~Hoffmann, P.~Gyugyi, H.~Wong-Toi, and G.~Franklin, ``Supervisory
  control of a rapid thermal multiprocessor,'' \emph{IEEE Transactions on
  Automatic Control}, vol.~38, no.~7, pp. 1040--1059, 1993.

\bibitem{BrandinCharbonnier:1994}
B.~Brandin and F.~Charbonnier, ``The supervisory control of the automated
  manufacturing system of the aip,'' in \emph{Proc. Rensselaer's 4th Int. Conf.
  Computer Integrated Manufacturing and Automation Technology}, 1994, pp.
  319--324.

\bibitem{Malik:2003}
P.~Malik, ``From supervisory control to nonblocking controllers for discrete
  event systems,'' Ph.D. dissertation, University of Kaiserslautern, 2003.

\bibitem{Elienberg:1974}
S.~Eilenberg, \emph{Automata, Languages and Machines Voluma A}.\hskip 1em plus
  0.5em minus 0.4em\relax Academic Press, 1974.

\bibitem{ZhangWangCai:2021}
R.~Zhang, Z.~Wang, and K.~Cai, ``N-step nonblocking supervisory control of
  discrete-event systems,'' in \emph{Proc. 2021 60th \text{IEEE} Conference on
  Decision and Control (CDC)}, Austin, Texas, December 13-15 2021, pp.
  339--344.

\bibitem{ZhangCaiAutomatica:2024}
R.~Zhang, J.~Wang, Z.~Wang, and K.~Cai, ``Quantitatively nonblocking
  supervisory control of discrete-event systems,'' \emph{Automatica}, 2024,
  accepted (Full version is available at {https://arxiv.org/abs/2108.00721}).

\bibitem{BerardEt:2001}
B.~Berard, M.~Bidoit, A.~Finkel, F.~Laroussinie, A.~Petit, L.~Petrucci, and
  P.~Schnoebelen, \emph{Systems and Software Verification}.\hskip 1em plus
  0.5em minus 0.4em\relax Springer-Verlag Berlin Heidelberg, 2001.

\bibitem{BonakdarpourKulkarni:2006}
B.~Bonakdarpour and S.~Kulkarni, ``Complexity issues in automated addition of
  time-bounded liveness properties,'' Department of Computer Science and
  Engineering, Michigan State University, Tech. Rep., 2006.

\bibitem{Kronos:2002}
C.~Daws, A.~Olivero, S.~Tripakis, and SergioYovine, \emph{Kronos: A
  verification tool for real-time systems}, 2002, available at
  \url{https://www-verimag.imag.fr/DIST-TOOLS/TEMPO/kronos/}.

\bibitem{QueirozEt:2005}
M.~H. de~Queiroz, J.~Cury, and W.~Wonham, ``Multitasking supervisory control of
  discrete-event systems,'' \emph{Discrete Event Dynamic Systems}, vol.~15,
  no.~4, pp. 375--395, 2005.

\end{thebibliography}

\end{document}